\documentclass[12pt]{spieman}  
\usepackage{amsmath,amsfonts,amssymb}
\usepackage{graphicx}
\usepackage{setspace}
\usepackage{tocloft}
\usepackage{tabulary}
\usepackage{aas_macros}
\usepackage{mathtools}

\title{Simulations of a High-Contrast Single-Mode Fiber Coronagraphic Multi-Object Spectrograph for Future Space Telescopes}

\author[a, b, *]{Carl T. Coker}
\author[b]{Stuart B. Shaklan}
\author[b]{A J E. Riggs}
\author[b]{Garreth Ruane}
\affil[a]{NASA Postdoctoral Program Fellow}
\affil[b]{Jet Propulsion Laboratory, California Institute of Technology, 4800 Oak Grove Drive, Pasadena, CA, USA, 91109}

\cftpagenumbersoff{figure}
\cftpagenumbersoff{table} 
\begin{document} 
\maketitle

\begin{abstract}
Directly imaging and characterizing Earth-like exoplanets is a tremendously difficult instrumental challenge.  Present coronagraphic systems have yet to achieve the required $10^{-10}$ broadband contrast in a laboratory environment, but promising progress towards this goal continues.  A new approach to starlight suppression is the use of a single-mode fiber behind a coronagraph.  By using deformable mirrors to create a mismatch between incoming starlight and the fiber mode, a single-mode fiber can be turned into an integral part of the starlight suppression system.  In this paper, we present simulation results of a system with five single-mode fibers coupled to shaped pupil and vortex coronagraphs.  We investigate the properties of the system, including its spectral bandwidth, throughput, and sensitivity to low-order aberrations.  We also compare the performance of the single-mode fiber configuration with conventional imaging and multi-object modes, finding improved spectral bandwidth, raw contrast, background-limited SNR, and demonstrate a wavefront control algorithm which is robust to tip/tilt errors.
\end{abstract}

\keywords{coronagraphy, single-mode fibers, wavefront control, exoplanets}

{\noindent \footnotesize\textbf{*}Carl T. Coker,  \linkable{carl.t.coker@jpl.nasa.gov} }

\begin{spacing}{2}


\section{Introduction}\label{sec:intro}

The detection and characterization of Earth-like exoplanets is fast becoming one of the preeminent problems in exoplanet science.  Probing these worlds would allow us to answer fundamental questions about the universe and our place in it, including whether we are alone.  However, actually doing so is incredibly difficult, including via the direct imaging technique.  Very high planet-to-star contrast ratios ($\sim10^{-10}$) at small angular separations ($0.1-1''$) are required to resolve these planets, and their inherent faintness, as well as the presence of backgrounds such as exozodi, necessitates long exposures, with spectral characterization times stretching into the hundreds of hours\cite{Wang17}.

One way to provide the required contrast is with a coronagraph.  Coronagraphs seek to internally block the light from the central star and reveal the faint planets and other targets which would ordinarily be overwhelmed.  There is a profusion of coronagraph types\cite{Soummer05,Foo05,Mawet05,Kasdin03,Guyon05,Kuchner02,Trauger11}, many of which have achieved $10^{-10}$ contrast or better in simulations\cite{Coker18,Ruane18,NDiaye16,Ruane16,Zimmerman16,Fogarty18}, and efforts are underway to demonstrate this contrast in the laboratory, with several having bettered $10^{-9}$ contrast\cite{Kern13,Serabyn14,Trauger07,Trauger11}.

One potential way to help overcome the $10^{-10}$ contrast barrier is to use one or more single-mode fibers (SMFs) to take light from the image plane to a characterization instrument such as a high-resolution spectrometer\cite{Snellen15}.  SMFs provide for very strong rejection of most light because they only guide one mode - if the shape of the incoming wavefront does not match the fiber mode, it is rejected by the fiber\cite{Shaklan88}.  In this way, it is possible to offload some of the starlight suppression from the coronagraph masks and other wavefront control elements onto the fiber itself.  This may allow for improved throughput and spectral bandpass over conventional coronagraphic systems, as a trade may be made between a coronagraph mask's raw contrast and its throughput and bandpass for most coronagraph types (see, e.g., Ref.~\citenum{Coker18}).

Previous work on coupling SMFs with coronagraphs has been promising.  The Single-mode Complex Ampltiude Refinement (SCAR) coronagraph design\cite{SCAR1,SCAR2} uses a pupil-plane phase plate to mode-shape the incoming starlight so that it is rejected by the fibers, achieving contrast of better than $10^{-4}$ over a broad band ($\sim20\%$) with high total system throughput (up to $50\%$).  Ref.~\citenum{Mawet17} showed that traditional speckle nulling techniques are greatly enhanced by SMFs, demonstrating in the laboratory additional nulling of a factor of $10^3$ over the nominal starlight suppression level of the coronagraph in monochromatic light.  Ref.~\citenum{Sayson18} demonstrated the use of electric field conjugation (EFC)\cite{Give'on07} with a SMF, using it as an integral part of the wavefront control system of the coronagraph.  In addition to their demonstration in coronagraph image planes, SMFs have served as spatial filters in the pupil plane of a visible nulling coronagraph\cite{Sandhu15}, where their spatial filtering properties are useful in conjunction with a segmented-mirror wavefront control system\cite{Clampin04}.

In this paper, we present results of our simulations of a SMF-fed multi-object spectrograph, with a focus on future space telescope concepts such as HabEx\cite{Mennesson16} and LUVOIR\cite{Bolcar17}.  We integrate full EFC control through the fibers with shaped pupil and vortex coronagraph masks to investigate the total system performance, including maximum allowable bandpass, throughput, and low-order aberration sensitivity, as well as the effect of observing a finite-sized stellar disk.  We further compare these results to those of a traditional imaging system.

\section{Model Design and Parameters}\label{sec:model}

\begin{figure}
\centering
\includegraphics[scale=0.55]{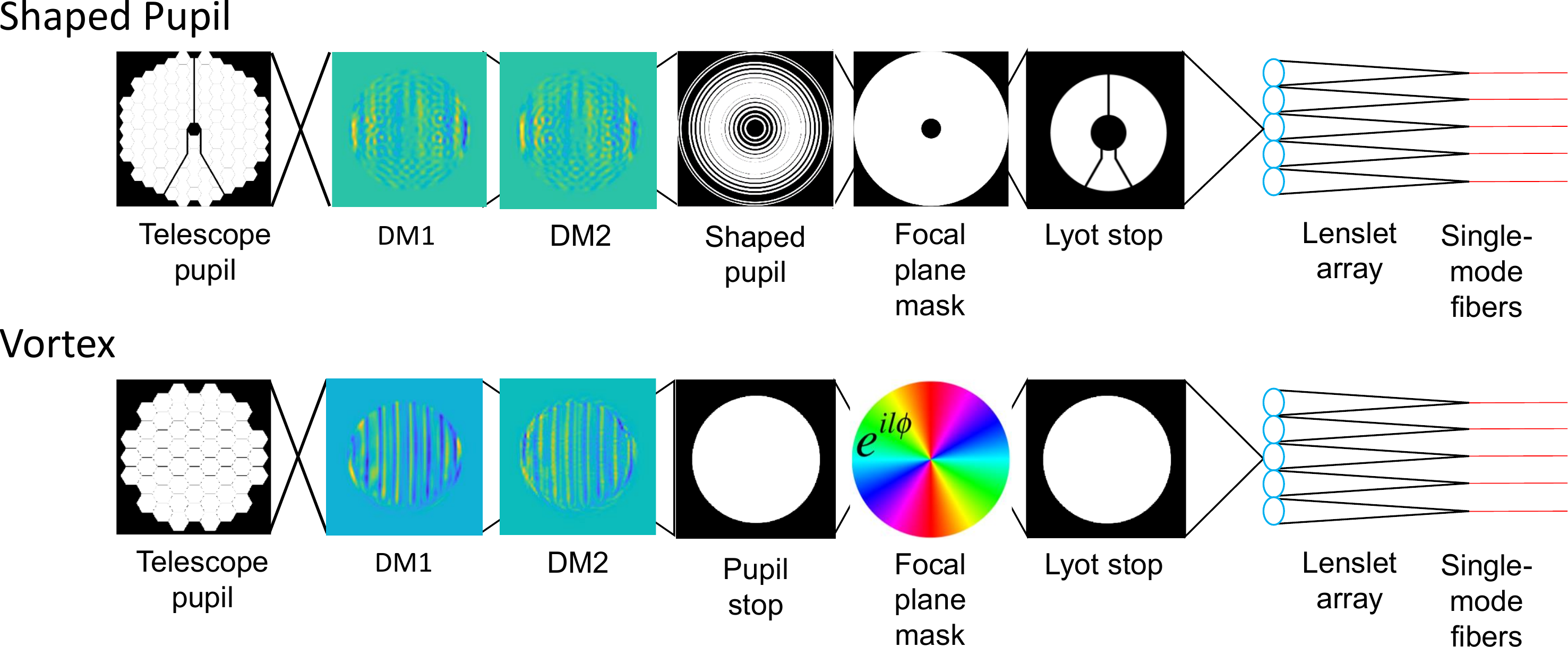}
\caption{Optical configurations for the shaped pupil Lyot (SPLC) and vortex architectures we tested.  The top chain is for the SPLC, the bottom is for the vortex.  The SPLC system uses the on-axis 15 m aperture LUVOIR A pupil, two DMs to control diffraction from the struts and segment gaps, a binary shaped pupil apodizer, a hard-edged focal plane mask, and a Lyot stop.  The vortex system uses the off-axis 8 m aperture LUVOIR B pupil, two DMs to control diffraction from the segment gaps, a pupil stop to circularize the edge of the pupil, an achromatic vortex phase mask, and a Lyot stop.  Both systems then feed into lenslets situated in the focal plane, which focus the light onto individual single-mode fibers.}\label{fig:opticaltrains}
\end{figure}

We modeled two different coronagraph architectures, one each for segmented on- and off-axis telescopes, each with two possible coronagraphic masks.  For the on-axis case, we used shaped pupil ring apodizers with hard-edged focal plane masks and Lyot stops, designed for lower ($10^{-8}$) and high ($10^{-10}$) contrast.  For the off-axis case, we used achromatic charge 6 and 8 vortex focal plane masks.  In both cases, we used two 64x64-actuator deformable mirrors (DMs) to control diffraction from the telescope struts (if any) and segment gaps.  We used the LUVOIR A and B pupils\cite{Bolcar17}, which are 15 and 8 meters in diameter, for the on- and off-axis cases, respectively.  We did not test the baseline HabEx pupil, as that is a completely unobscured circular aperture, and should thus offer the same as or better performance than either LUVOIR architecture.  Figure~\ref{fig:opticaltrains} shows the optical configuration for each architecture.  We did not include optical element surface errors or reflection losses.

To model the coronagraph/fiber system, we used the Fast Linearized Coronagraph Optimizer (FALCO)\cite{Riggs18}.  FALCO models the optical propagation through the telescope/coronagraph system and uses the EFC algorithm for wavefront control, varying the regularization parameter to find the DM shape which minimizes the intensity in the image plane by default.  Thanks to improvements in the linearized DM response matrix calculation, FALCO is generally much faster than previous DM-integrated coronagraph design codes and is capable of generating wavefront control solutions orders of magnitude faster than a brute-force approach.  For our purposes, we retrofitted a model of a lenslet array feeding into single-mode fibers onto the back end of the main FALCO code, additionally changing the EFC response matrix calculation to minimize the starlight in the fiber.  We simulated lenslets here because they allow for denser sampling of the image plane than individual SMFs place in the focal plane; see Section~\ref{sec:conclusion} for a more thorough discussion of this.  The fiber model defines the physical radius of the fiber core and its numerical aperture, from which we calculate the mode shape and overlap integral.  As a result, the model fully captures the coupling efficiency of the fiber and includes this in throughput calculations.

The fiber mode shape itself is generated using the assumption of a perfect step-index fiber with infinitely thick cladding, which results in the following electric field:\cite{Jeunhomme83}
\begin{equation}
\mathbf{E} = \mathbf{E}_0 \times 
\begin{dcases}
	\frac{J_0 (Ur/a)}{J_0 (U)} & 0 \leq r \leq a \\
	\frac{K_0 (Wr/a)}{K_0 (W)} & r \geq a
\end{dcases}
\end{equation}
where $r$ is the radial coordinate, $a$ is the radius of the fiber core, $J_0$ and $K_0$ are Bessel and modified Bessel functions of the first kind, respectively, and the constants $U$ and $W$ are related to fiber properties by
\begin{align}
V &= \frac{2\pi}{\lambda} a\mathrm{NA} \\
W &= 1.1428V - 0.996 \\
U &= \sqrt{V^2 - W^2}
\end{align}
where NA is the numerical aperture of the fiber.  We chose the fiber properties such that $V=2.405$ at 350~nm, so that the fiber is exclusively single-mode for the entirety of the tested bandpasses.

Table~\ref{table:coroparams} contains the design parameters for each architecture.  For the shaped pupil Lyot coronagraph (SPLC), we use coronagraph masks designed to reach $10^{-10}$ and $10^{-8}$ contrast (hereafter called SPLC 1 and 2, respectively) on annular pupils without struts or mirror segment gaps to investigate whether we could leverage the additional starlight suppression power of SMFs to improve performance in other areas.  For the vortex, we were more concerned with whether the additional aberration control of the charge 8 mask over the charge 6 was necessary to maintain acceptable contrast.

For each coronagraph architecture, we tested the maximum achievable bandpass ($\Delta\lambda/\lambda$) at $10^{-10}$ contrast, determined the throughput over the maximum usable band, and investigated the sensitivity of the wavefront solution to low-order aberration drifts.  We attempted to minimize the starlight coupling into five lenslet/fiber pairs simultaneously with each architecture; we used five lenslets as a mock for the largest number of planets that could feasibly be observed simultaneously in a single star system.  The lenslets were spaced evenly in a line stretching from the inner working angle (IWA) for each coronagraph; we also performed one trial with randomized lenslet positions, and performance did not degrade (Section~\ref{sec:conclusion}).  The lenslet radius in each case was $1.6\lambda_0/D$, where $\lambda_0$ is the central wavelength, designed to maximize coupling into the fiber for a source on the optical axis of the lenslet.

\begin{table}
\centering
\caption{Coronagraph Parameters}
\begin{tabulary}{\textwidth}{C|C|C|C|C|C}
\hline
\hline
Coronagraph Mask & Aperture Type & Imaging Mode Bandwidth ($\Delta\lambda/\lambda$) & Imaging Mode FWHM Throughput at $6\lambda_0/D$ & Inner Working Angle ($\lambda/D$) & Lenslet Positions ($\lambda_0/D$) \\ \hline
$10^{-10}$ SPLC (SPLC 1) & On-axis centrally obstructed segmented & 10\% & 0.12 & 4 & 4, 9, 14, 19, 24 \\ \hline
$10^{-8}$ SPLC (SPLC 2) & On-axis centrally obstructed segmented & 10\% & 0.18 & 4 & 4, 9, 14, 19, 24 \\ \hline
Charge 6 Vortex & Off-axis segmented & 20\% & 0.25 & 3 & 3, 8, 13, 18, 23 \\ \hline
Charge 8 Votex & Off-axis segmented & 20\% & 0.22 & 4 & 4, 9, 14, 19, 24\\
\hline
\end{tabulary}
\label{table:coroparams}
\end{table}

\section{Wavefront Control Using Single-Mode Fibers}\label{sec:WFC}

In standard EFC, the optical system is represented as a linear operator on the incoming light field, and the DM phase change is linearized to be entirely imaginary.  This then allows the DM shape that minimizes the intensity in the image plane to be solved with linear algebra, assuming the complex electric field in the image plane is known.  This is represented in Equation~4 of Ref.~\citenum{Give'on07}, which we reproduce here:
\begin{equation}
G\overline{a} = iE_{ab}
\end{equation}
where $G$ is the linearized response matrix, or Jacobian, for the DMs, $\overline{a}$ is the matrix of DM actuator lengths, and $E_{ab}$ is the electric field in the image plane due to the aberrations we wish to correct.  This can then be solved for $\overline{a}$ via left pseudo-inverse and regularization.  In order to control multiple wavelengths, more rows can be added to the matrix, representing the Jacobian and electric field values at each new wavelength; two DMs can be similarly accounted for by adding extra columns to the response matrix for the second DM.  Any number of point sources may be included in the Jacobian as well.  This property of EFC can be used to make the wavefront solution more robust against tip/tilt and the finite size of the on-axis star by modeling the star as a small collection of point sources.

SMFs add a new wrinkle to standard EFC.  Instead of minimizing the starlight in the final telescope focal plane, we can instead attempt to minimize the amount of starlight being coupled into a SMF (or multiple SMFs simultaneously).  In essence, we wish to minimize the fiber coupling efficiency for starlight, given by
\begin{equation}
\eta = \frac{\left|\int FM \cdot E^{\star}_{ab} dA\right|^2}{\int |FM|^2 dA \int |E^{\star}_{ab}|^2 dA}
\end{equation}
where FM is the fiber mode shape.  In this way, we ask the DM to only control the starlight which goes into the fibers, leaving the rest of the dark hole bright.  Compared to controlling a full dark hole, this approach has several benefits:  first, it reduces the number of points in the Jacobian matrix from twice the number of dark hole pixels (for real and imaginary components of E-field at each pixel) to just the number of fibers controlled, speeding computation time; second, it reduces the DM stroke required by at least a factor of 3-4 over digging a large dark hole because less area in the image is being controlled; and third, it allows us to leverage the increased starlight suppression capability of the SMF to improve coronagraph performance, as we will show in Sections~\ref{sec:imgresults}-\ref{sec:Vortexresults}.  For a full mathematical treatment of fiber EFC, see Ref.~\citenum{Sayson19}.

\begin{figure}
\centering
\includegraphics[scale=0.65]{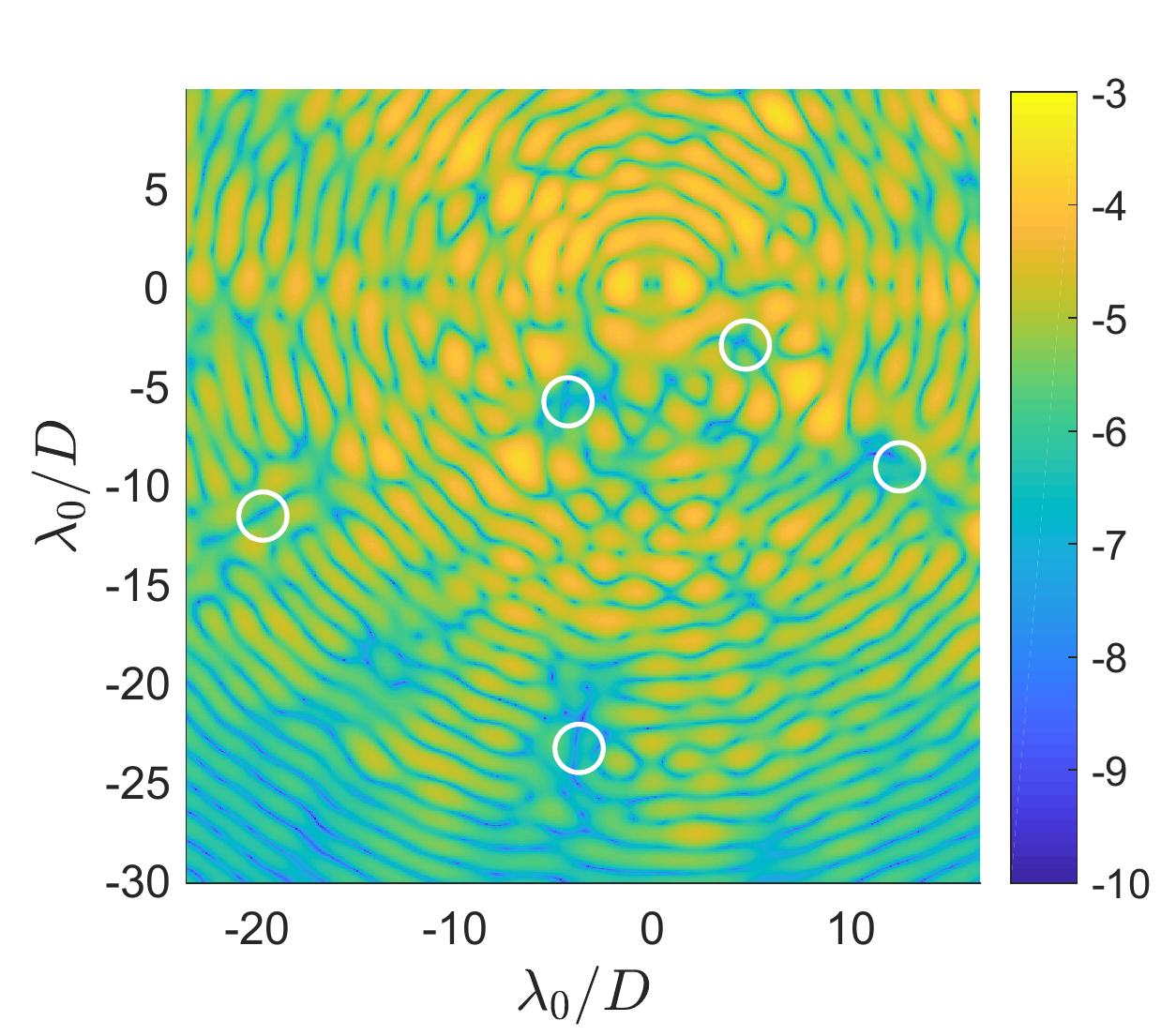}
\includegraphics[scale=0.65]{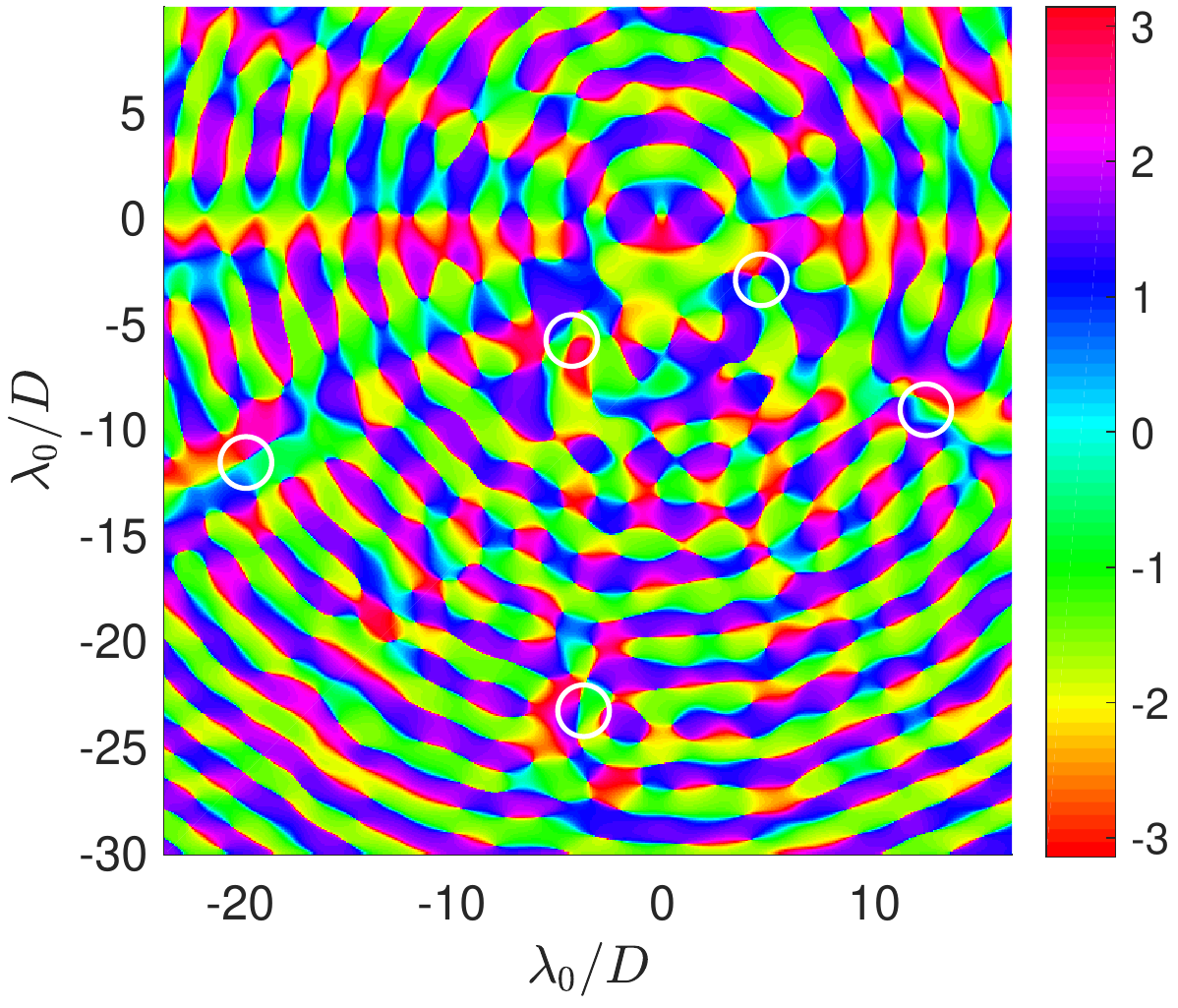}
\caption{Base-10 log normalized intensity and phase in the coronagraph image plane for a trial using SPLC 2 and randomized lenslet positions.  The left panel shows the base-10 log of the normalized intensity at the central wavelength of the band, while the right panel shows the phase at the same wavelength.  The lenslets are marked by the white circles.  The plots have been zoomed in to better show the phase patterns across the lenslets.  The coronagraph's dark hole remains quite bright, as do the regions atop the lenslets.  Instead, the phase structure of the field on the lenslet results in a mismatch with the fiber mode, leading to strong suppression of the starlight in the fibers.}\label{fig:SMFintphase}
\end{figure}

Figure~\ref{fig:SMFintphase} shows the normalized intensity and phase structure in the telescope focal plane for one of our trials using randomized lenslet positions (the setup of this trial is described in more detail in Section~\ref{sec:conclusion}).  Note how the bright the dark hole remains, even atop the fibers themselves, ranging between $10^{-5} - 10^{-7}$ raw contrast on average.  Instead of canceling out most of the light in the focal plane, the solution relies on using the phase structure to reject most of the starlight hitting the fiber, resulting in a deep null after the SMF.  For point source control such as in Figure~\ref{fig:SMFintphase}, FALCO using EFC mostly finds first-order nulls across the lenslets (i.e., there is only one phase flip across a lenslet), although second-order nulls  and other shapes such as vortex phase ramps are possible depending on the existing phase pattern in the focal plane, or whether tip/tilt sensitivity control is included in the EFC Jacobian.  Despite there being a relatively large amount of light falling onto the fibers, the aberration sensitivity is not appreciably increased near the IWA (Sections~\ref{sec:SPLCresults}-\ref{sec:Vortexresults}).

\section{Initial Imaging Results}\label{sec:imgresults}

\begin{figure}
\centering
\includegraphics[scale=0.56]{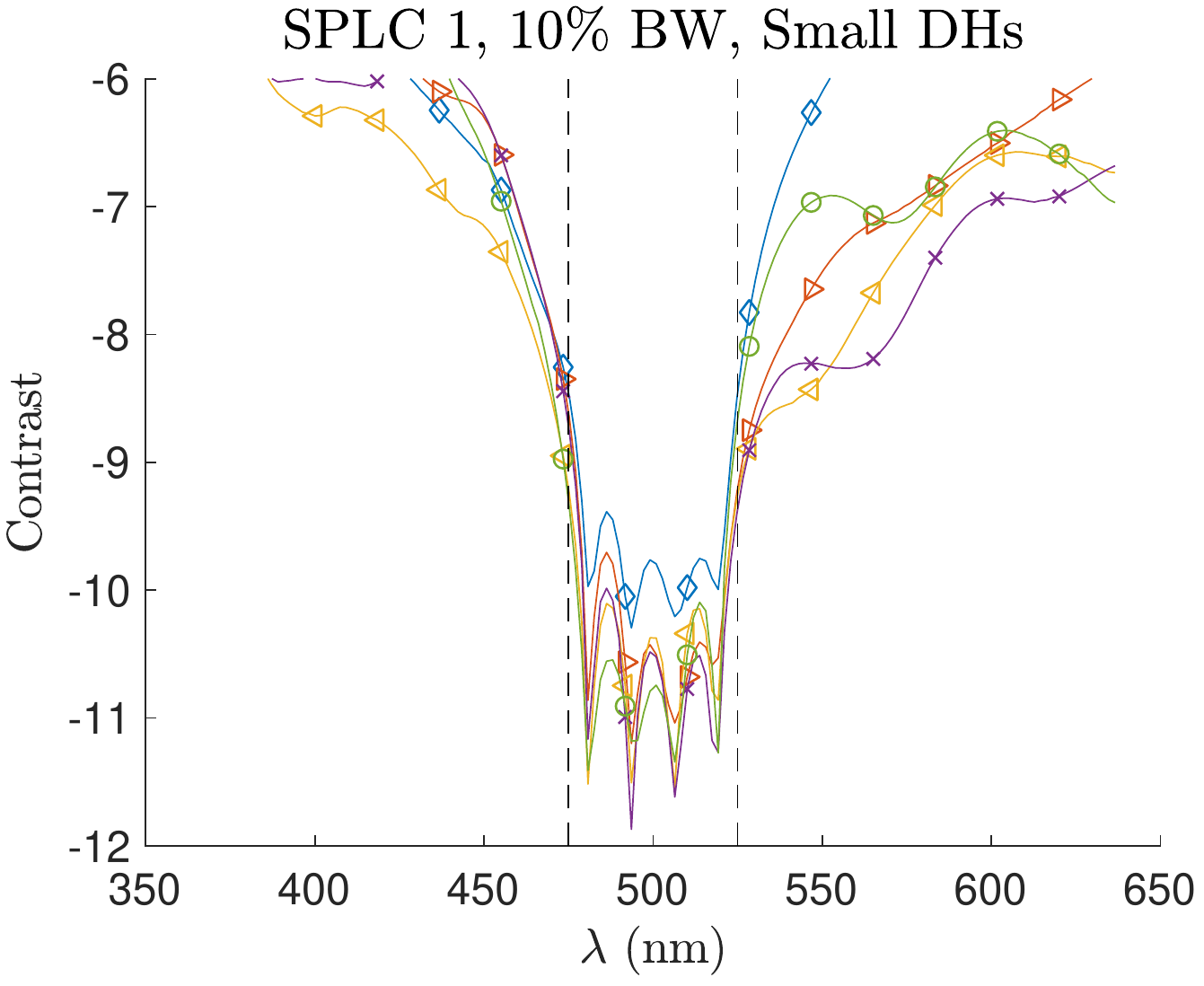}
\includegraphics[scale=0.56]{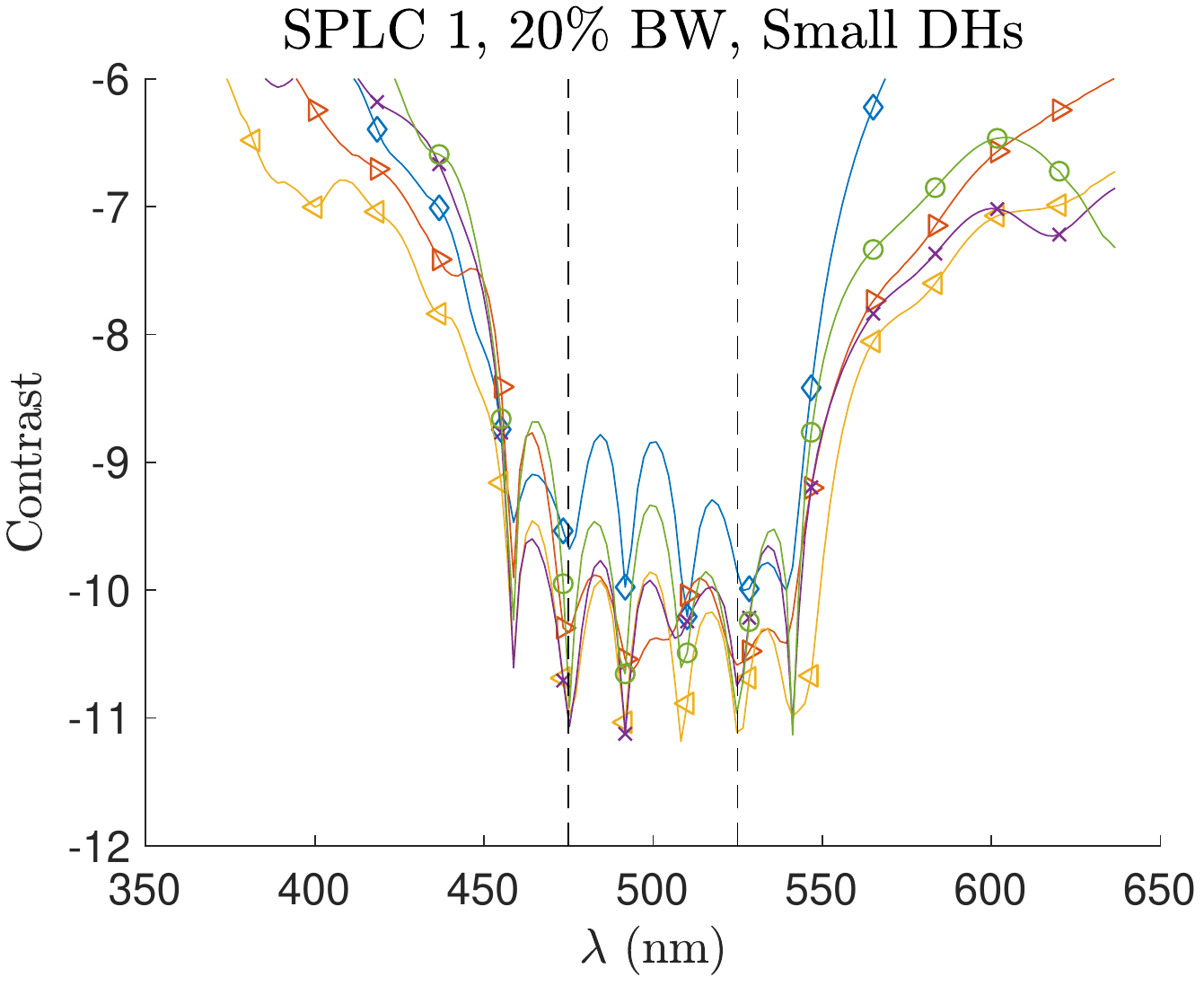}
\includegraphics[scale=0.56]{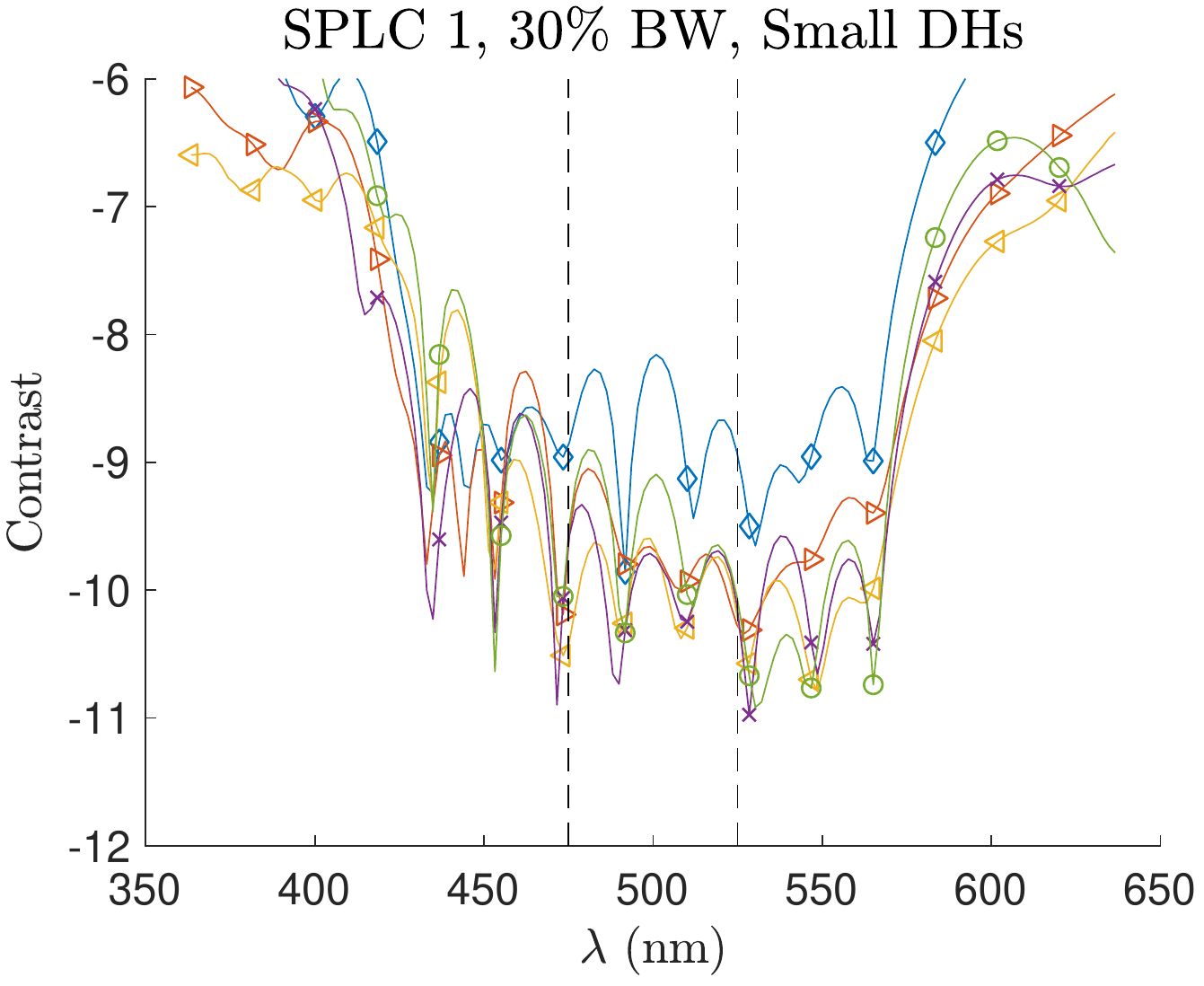}
\caption{Base-10 log raw contrast vs. wavelength for SPLC 1 in imaging mode, only trying to dig a dark hole on top of the five lenslet locations.  Each line represents the mean raw contrast over one lenslet; the blue diamond-marked line is the lenslet at the IWA.  The upper left panel shows the results for a 10\% bandwidth simulation, the upper right panel shows a 20\% bandwidth model, and the lower center panel shows a 30\% bandwidth model.  Note that while the 10\% bandpass dark holes are deeper than $10^{-10}$ further out into the field, none of the 30\% bandwidth dark holes achieve that contrast.}\label{fig:SPLCimgBW}
\end{figure}

In order to set a baseline of comparison for the SMF simulations, we also ran a series of simulations of SPLC 1 and the charge 6 vortex mask in imaging mode, attempting to make dark holes only over the top of the lenslets, assuming a camera is placed in the lenslet plane.  Figure~\ref{fig:SPLCimgBW} shows the contrast vs. wavelength for the SPLC imaging mode simulations using 10\%, 20\%, and 30\% bandwidth.  Depending on location in the field, SPLC 1 is able to achieve as low as $2\times10^{-11}$ in raw contrast in the 10\% bandwidth solution, showing that there is a benefit to focusing the dark hole on a small region of the field if the position of the target is known.  As the bandwidth is increased, the performance of the coronagraph decreases following the expected $C \propto \Delta\lambda^2$ until at 30\% bandwidth, $10^{-10}$ contrast is no longer achievable except in a restricted portion of the bandpass near the outer working angle.

\begin{figure}
\centering
\includegraphics[scale=0.56]{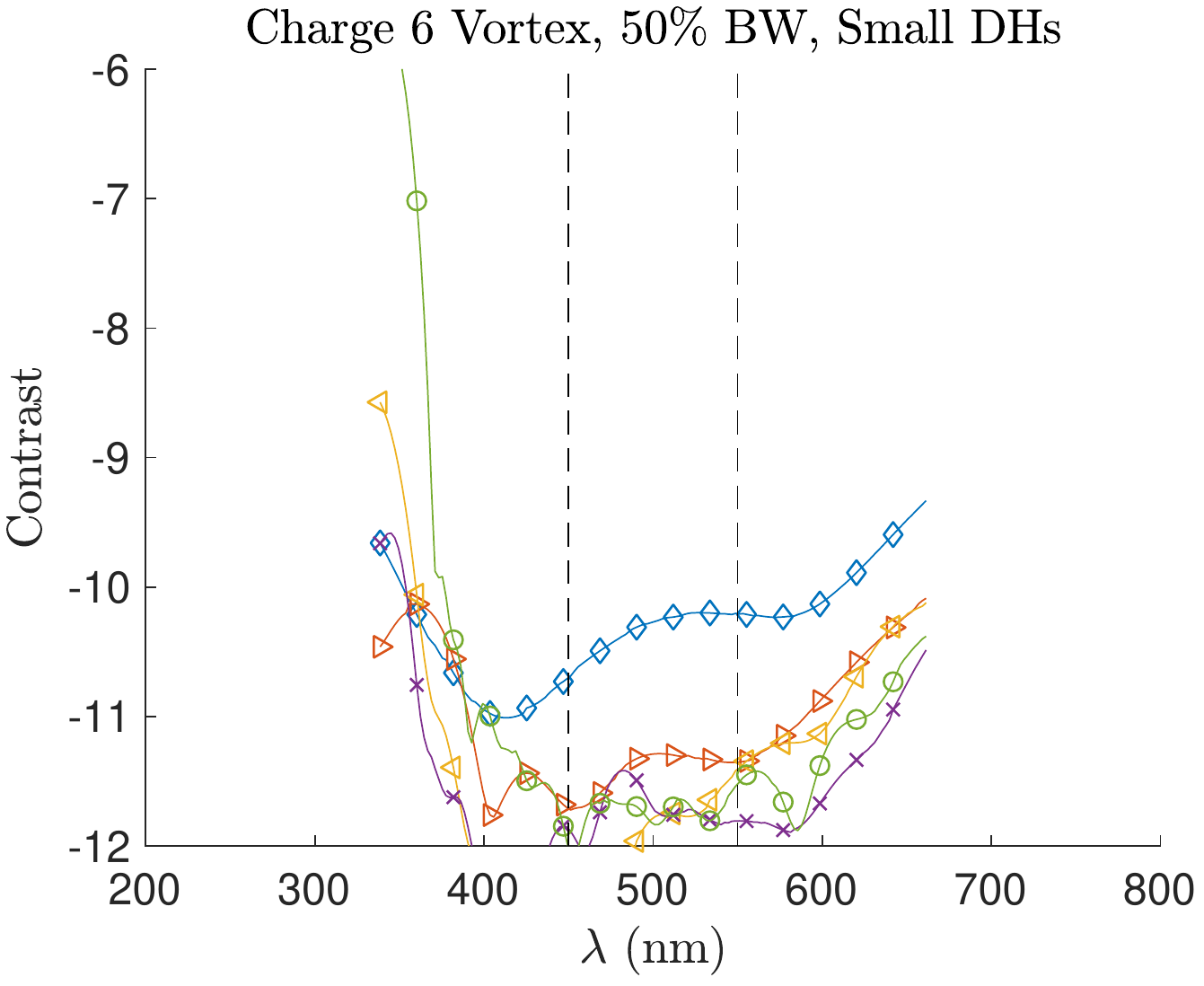}
\includegraphics[scale=0.56]{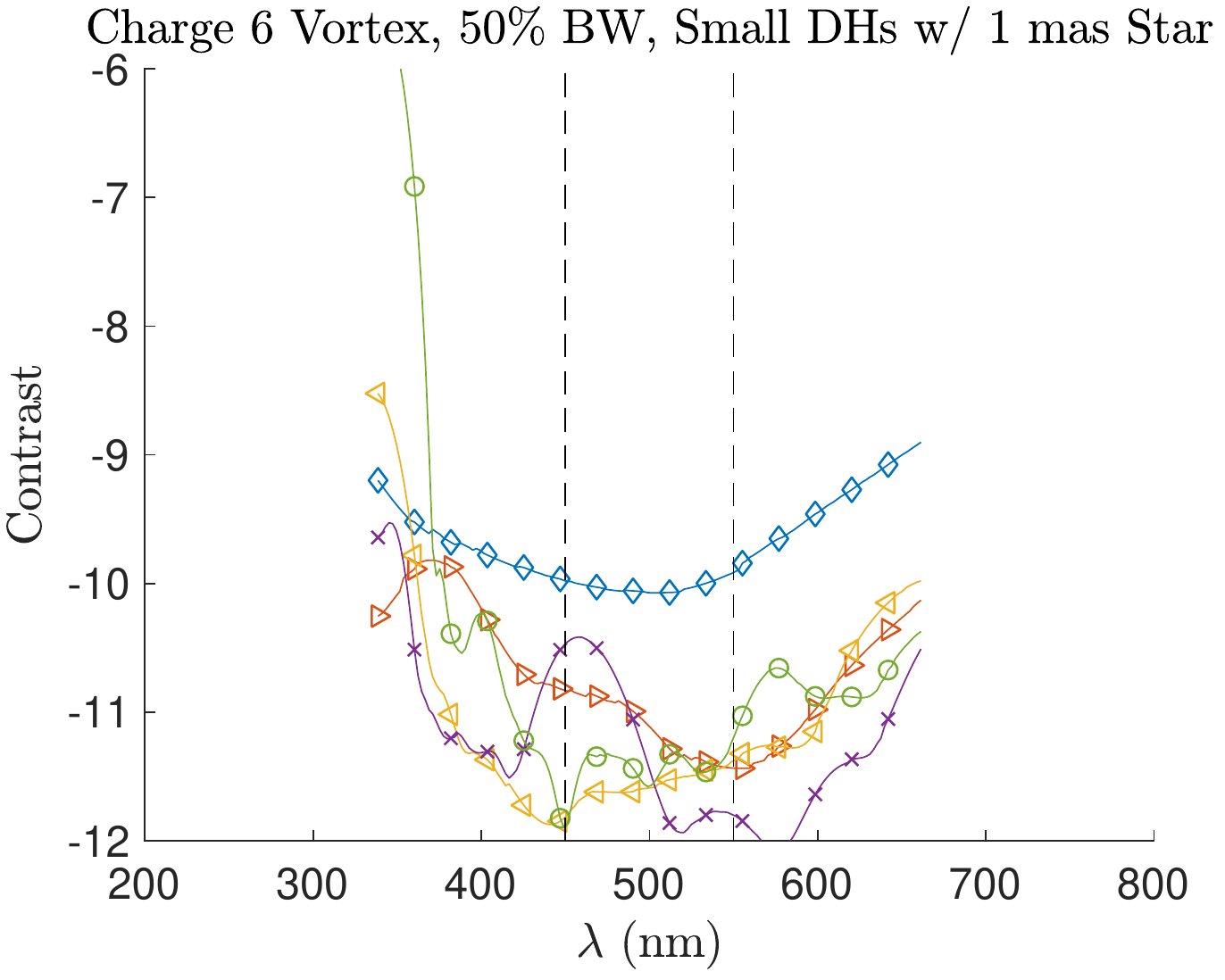}
\caption{Base-10 log raw contrast vs. wavelength for the charge 6 vortex mask in imaging mode, making a dark hole only at the locations of the five lenslets.  The left panel shows results for a point source, while the right panel uses a 1~mas diameter star.  Each line represents one lenslet; the blue diamond-marked line is the lenslet at the IWA, while the green line is furthest from the center of the focal plane.  The vertical dashed black lines show the nominally expected 20\% bandwidth for imaging mode.  Unlike the SPLC, a truly achromatic vortex mask allows for suppression over very broad bandpasses.}\label{fig:VorteximgBW}
\end{figure}

Figure~\ref{fig:VorteximgBW} shows the contrast vs. wavelength for the charge 6 vortex mask in imaging mode with a point-like and a finite-sized star when digging a dark hole solely over the locations of the five lenslets.  Each lenslet achieved $10^{-10}$ contrast or better over at least a 50\% band for the point source, while the finite-star results show similar performance between the SMF and limited dark hole imaging case.  The furthest lenslet out fails blueward of 375~nm due to exceeding the $32\lambda/D$ control radius of our DMs at those wavelengths.  We believe that the reason for such high performance in imaging mode is the perfect achromaticity assumption for the vortex phase mask in our simulations; vortex coronagraphs have not been demonstrated to achieve deep contrast beyond 20\% bandwidth\cite{Serabyn16}.  Imperfections in the mask and dispersion will be introduced in future models.

\section{DM-Augmented Shaped Pupil Lyot coronagraph \& SMF Results}\label{sec:SPLCresults}

\subsection{Maximum Bandpass and Finite-Sized Star}

\begin{figure}
\centering
\includegraphics[scale=0.572]{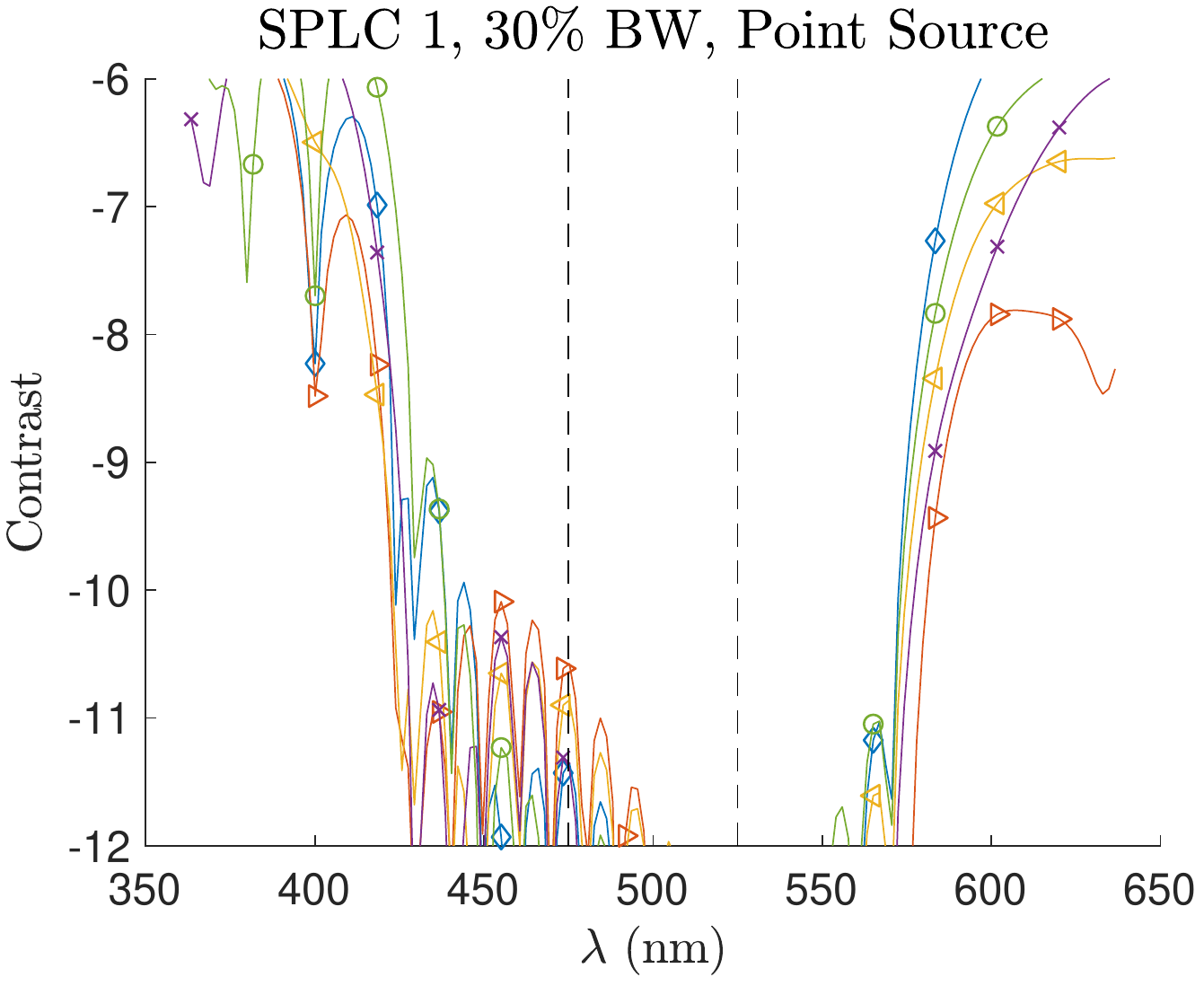}
\includegraphics[scale=0.572]{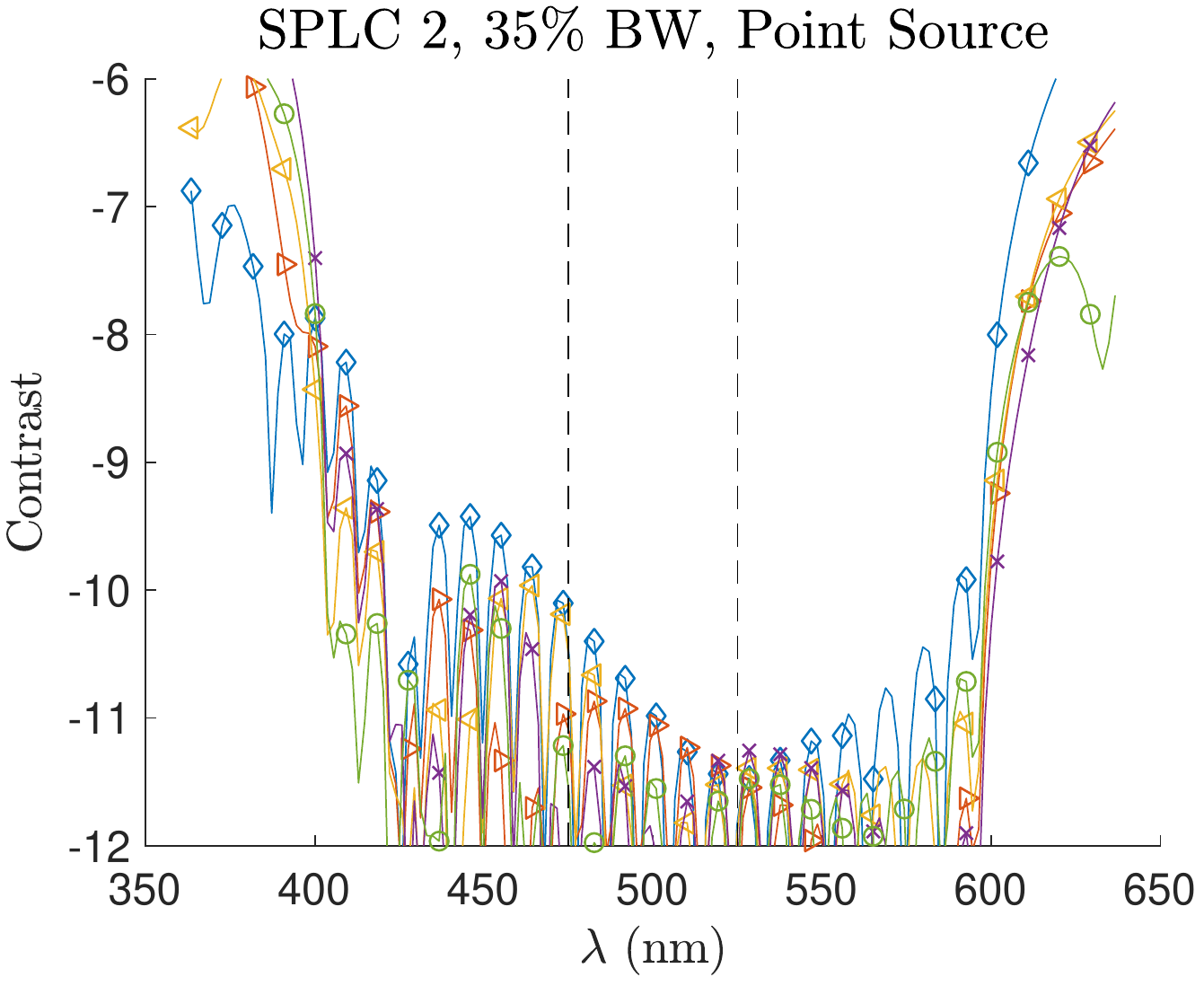}
\includegraphics[scale=0.572]{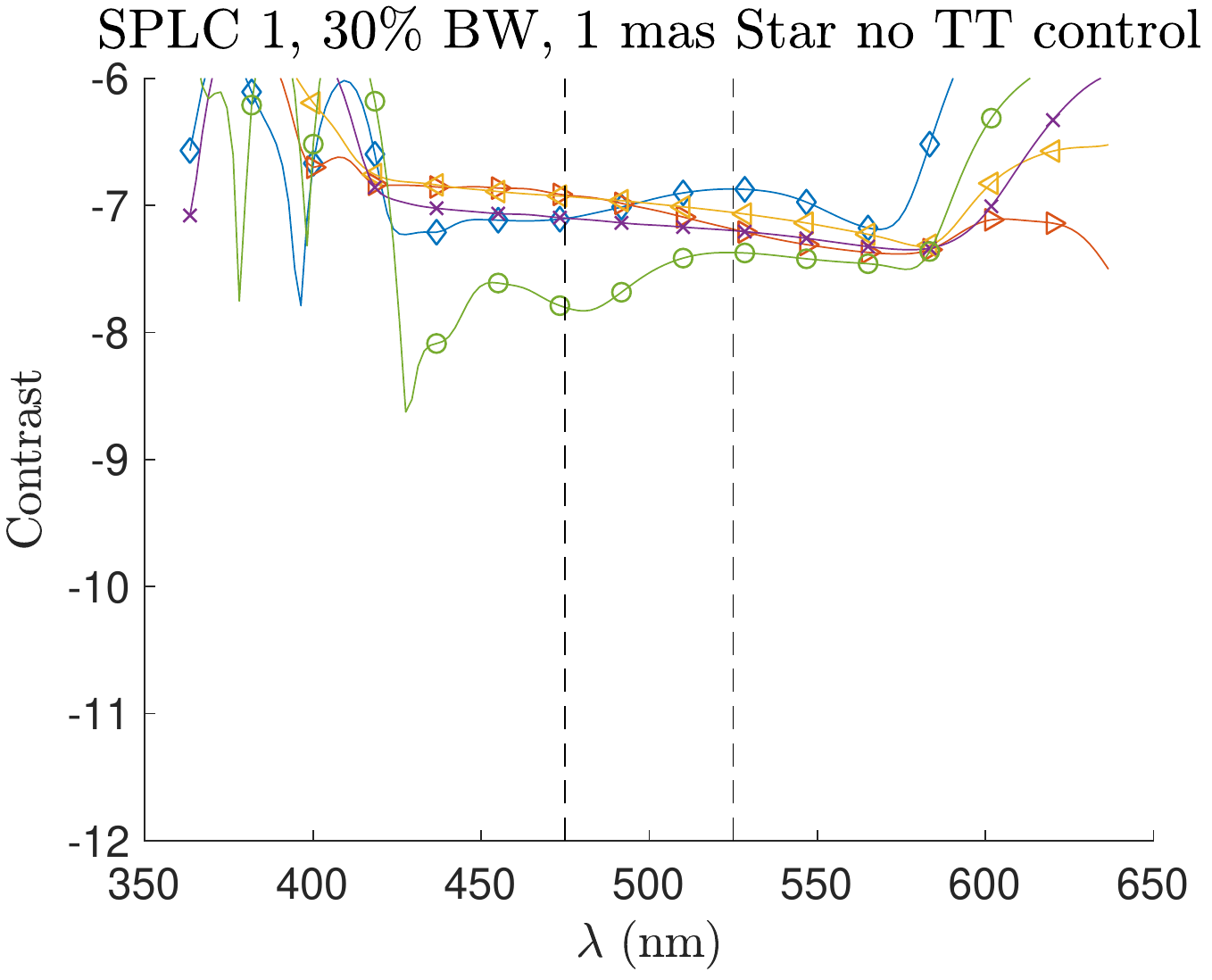}
\includegraphics[scale=0.572]{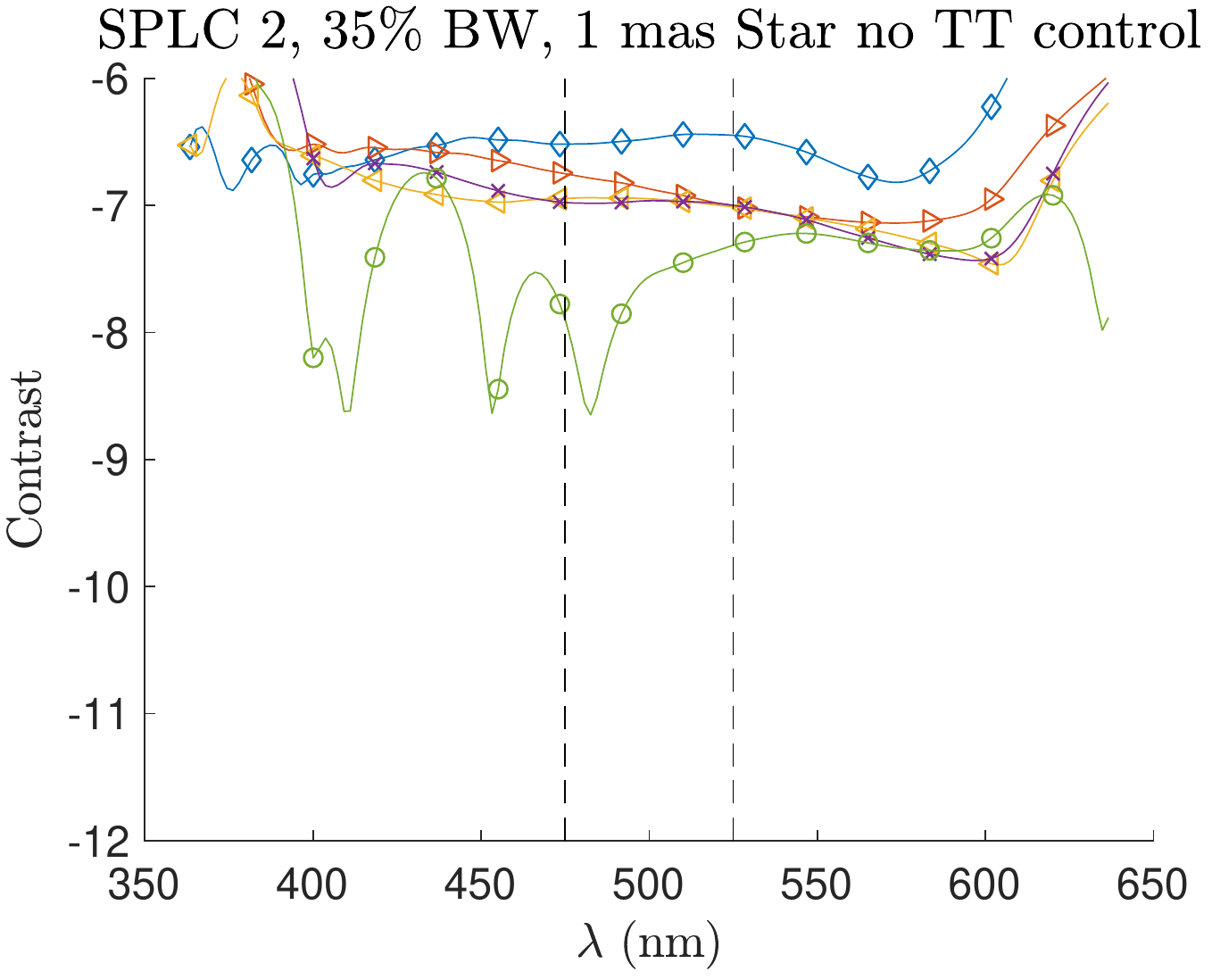}
\includegraphics[scale=0.572]{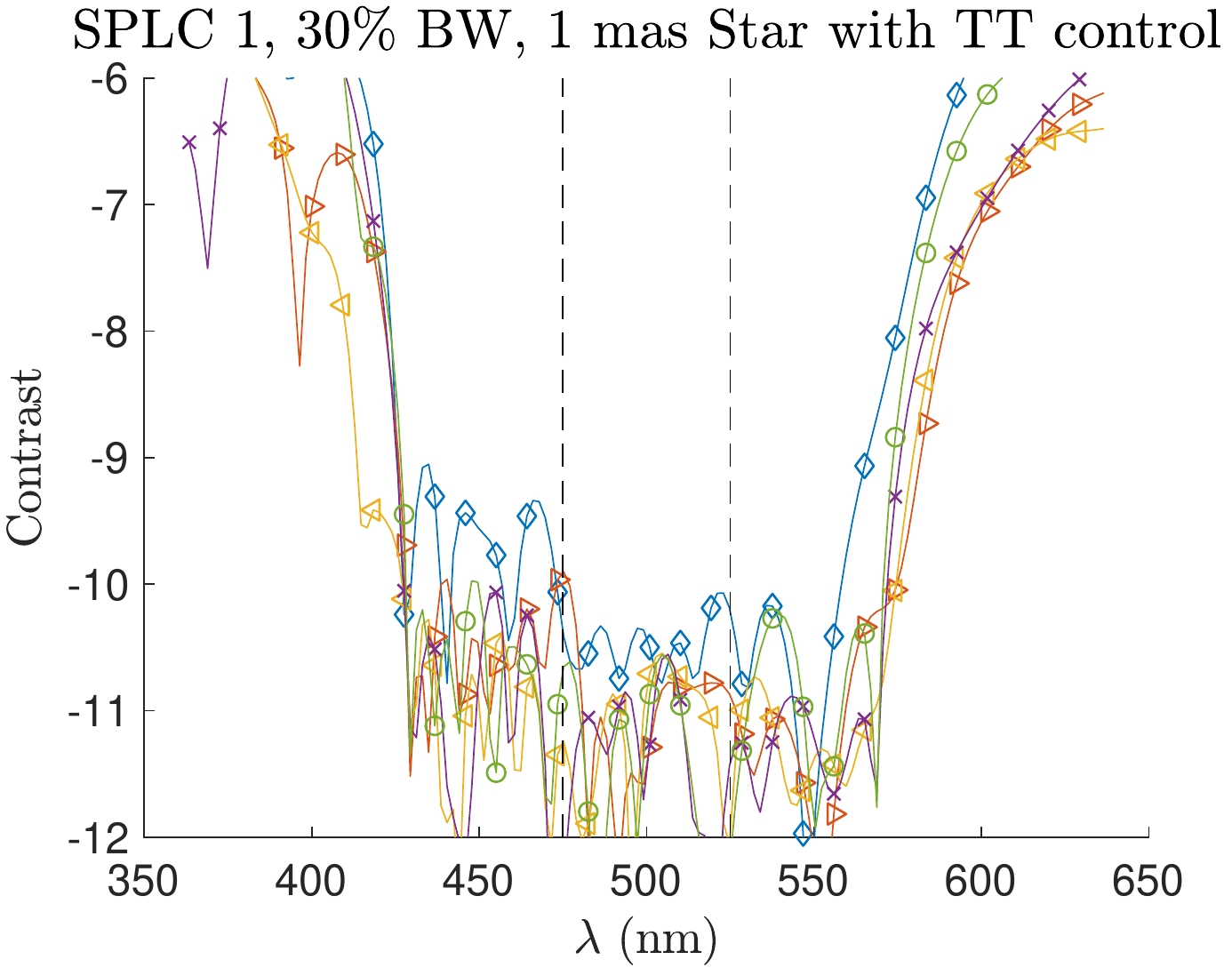}
\includegraphics[scale=0.572]{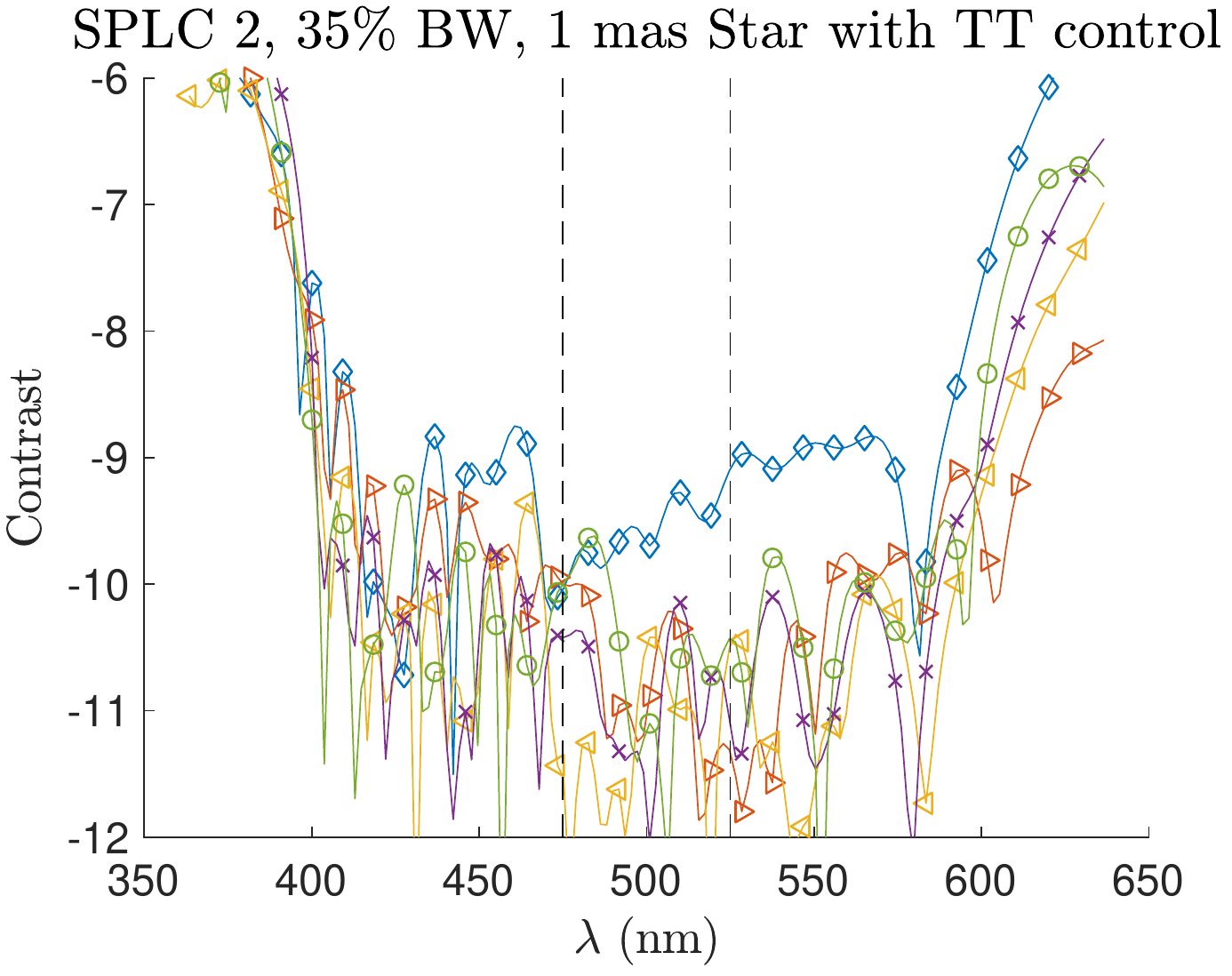}
\caption{Achieved base-10 log raw contrast vs. wavelength for each fiber in the SPLC mask cases.  The left column shows the results for SPLC 1, while the right column shows the results for SPLC 2.  The blue diamond-marked line in each panel shows the fiber at the IWA, while the vertical dashed black lines show the original design bandwidth of the coronagraph masks (in this case 10\%).  The top panels show the results from a point source, the middle panels the introduction of a 1~mas star using the point source wavefront solution, and the bottom panels the results using a 1~mas star with tip/tilt sensitivity control enabled.  The $10^{-10}$ mask achieves a 30\% bandwidth, while the $10^{-8}$ mask achieves a 35\% bandwidth, tremendously increasing the observational efficiency of the instrument.  Note also that with SMFs, many wavelengths must be controlled simultaneously, as the nulls are extremely chromatic.}\label{fig:SPLCBW}
\end{figure}

Single-mode fibers allow wavefront correction over a far larger wavelength range than coronagraph mask imaging modes.  The top row of Figure~\ref{fig:SPLCBW} shows achieved contrast as a function of wavelength for the maximum bandwidth solutions for the SPLC masks.  SPLC 1 can achieve the design contrast ($10^{-10}$) over a 30\% bandpass when using five fibers, while SPLC 2 achieves $10^{-10}$ contrast over a 35\% bandpass using the same number of fibers.  This represents a large efficiency improvement compared to the coronagraph's imaging mode, allowing a wide spectrum of an exoplanet to be taken in as little as a third of the time due to the increase in bandwidth.  Of particular note is that SPLC 2 was able to maintain $10^{-10}$ contrast across the entire 35\% band through the SMFs despite not being able to achieve better than $10^{-8}$ contrast in imaging mode.

It is typical in coronagraph simulations to treat the on-axis star as a point source to save computation time, but this is not realistic.  At the large telescope diameters that will be required to observe Earth-like exoplanets, the finite size of the star is very significant.  For example, on the LUVOIR A aperture that we used for our SPLC simulations, a 1~mas star represents $0.15\lambda/D$ at 500~nm wavelength.  This corresponds to a $\sim35$~nm RMS tip/tilt error, 350 times larger than the 100~pm RMS tip/tilt we consider later in Sections~\ref{sec:SPLCab} and \ref{sec:vortexab}.  The middle row of Figure~\ref{fig:SPLCBW} shows the contrast vs. wavelength for the SPLC simulations when the source is a 1~mas diameter star and no attempt at tip/tilt sensitivity control is made.  The effect is dramatic - on average, no better than $10^{-7}$ contrast is achieved across the band with either SPLC mask.

The situation is greatly improved when tip/tilt sensitivity control is included in the wavefront correction.  The bottom row of Figure~\ref{fig:SPLCBW} shows contrast vs. wavelength for the two SPLC masks with tip/tilt sensitivity control for a 1~mas star included in the wavefront correction Jacobian via offset sources being suppressed in parallel.  Although SPLC 1 is able to retain $10^{-10}$ contrast over the entire 30\% bandwidth, SPLC 2 is only able to maintain a consistent $10^{-9}$ at the IWA, though it nears or surpasses $10^{-10}$ further out in the field.  These results show the efficacy of tip/tilt sensitivity control when used for a small portion of the dark hole (previous work\cite{Coker18} showed that DM tip/tilt sensitivity control is ineffective for SPLCs when implemented over the entire controllable dark hole).

\subsection{Throughput and SNR}

\begin{figure}
\centering
\includegraphics[scale=0.56]{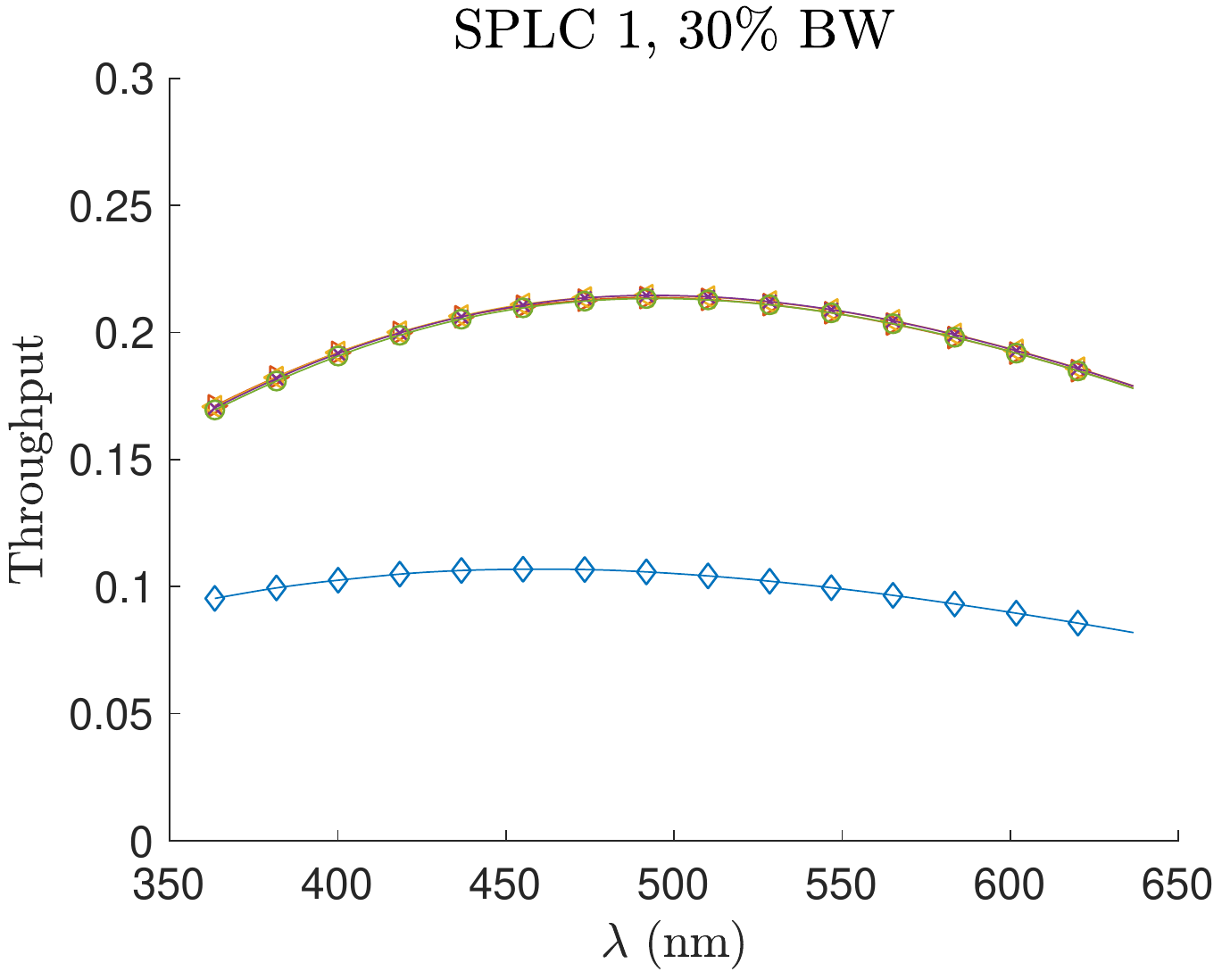}
\includegraphics[scale=0.56]{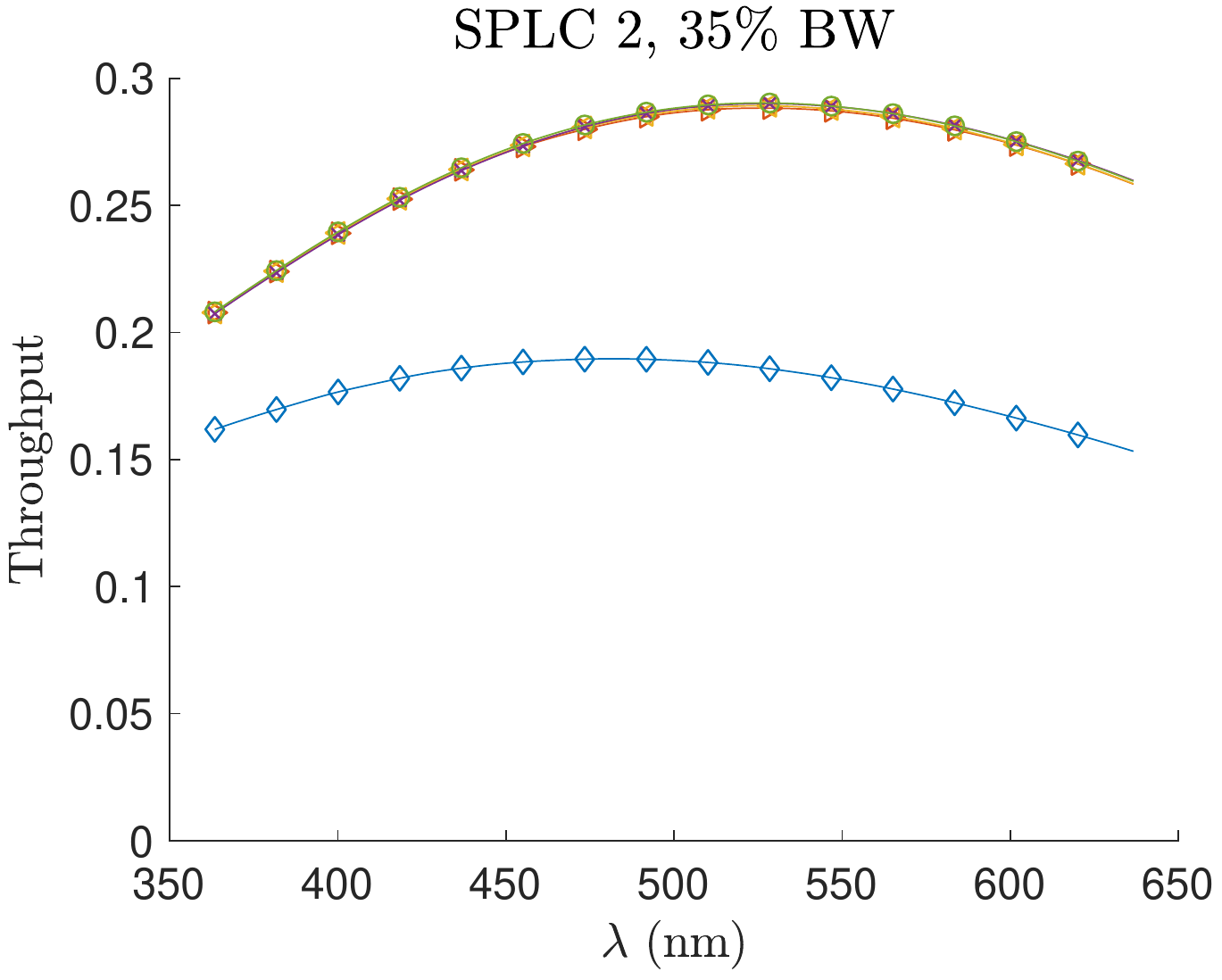}
\caption{Fiber throughput vs. wavelength for each fiber behind the SPLC masks.  The left panel shows the results for SPLC 1, while the right panel shows the SPLC 2 results.  Each line represents one fiber; the blue diamond-marked line in each panel corresponds to the IWA fiber.  Using SPLC 2 results in a 35\% increase in throughput over SPLC 1, while maintaining $10^{-10}$ contrast over a wider bandwidth than SPLC 1.}\label{fig:SPLCthput}
\end{figure}

The trade space available to coronagraph designers is illustrated in the difference in performance of a mask designed for $10^{-10}$ contrast (SPLC 1), and one designed for $10^{-8}$ contrast (SPLC 2), but with higher throughput and a slightly smaller inner working angle. Figure~\ref{fig:SPLCthput} shows throughput vs. wavelength for each fiber using our two SPLC masks.  When measuring fiber throughput, we took the ratio of all the light coming through the fiber relative to the light at the initial telescope pupil to generate a total system throughput.  This means that our throughput numbers are enhanced relative to the ``FWHM throughput'' in Table~\ref{table:coroparams}, as those numbers are measured from the FWHM of the PSF; the throughput we measure at the output of the fiber for SPLC 1 is exactly as expected when taking into account the encircled energy ratio ($\sim0.86/0.5$ comparing a full Airy pattern core to the FWHM) and fiber coupling efficiency.  Regardless, SPLC 2 shows a factor of $\sim1.4$ improvement in throughput over SPLC 1, including at the masks' respective IWAs.

A particular concern for exoEarth observations is the relatively uniform backgrounds which may be present at $10^{-10}$ contrast, whether from residual uncorrected starlight or from exozodiacal dust.  The average fiber throughput over a $1.22\lambda/D$ radius lenslet of a uniform background using SPLC 1 is 7.8\% at $6\lambda_0/D$ separation.  In imaging mode, the throughput of an infinite, uniform background is controlled strictly by the raw amount of light blocked by the apodizer mask, focal plane mask, and Lyot stop, as well as the aperture size used in the focal plane for photometry.  The relative background-limited SNR is $T_{\textrm{pl}}/\sqrt{T_{\textrm{bg}}D_{\textrm{ap}}^2}$, where $T_{\textrm{pl}}$ is the planet throughput, $T_{\textrm{bg}}$ is the average background throughput over the photometric aperture, and $D_{\textrm{ap}}$ is the diameter of the photometric aperture.  For the fiber case, $T_{\textrm{pl}} = 19\%$, $T_{\textrm{bg}} = 7.8\%$, and $D_{\textrm{ap}} = 2.44\lambda_0/D$, while for the imaging mode case, $T_{\textrm{pl}} = 12\%$, $T_{\textrm{bg}} = 41\%$ (simply the percentage of the telescope pupil light that makes it through the shaped pupil mask and Lyot stop), and $D_{\textrm{ap}} = \lambda_0/D$ (the FWHM of an Airy pattern).  This results in a factor of $\sim1.5$ increase in background-limited SNR in the same integration time by moving to SMFs from imaging.  The ultimate reason for this is that the coupling of PSFs that are misaligned with the SMF is very poor, and thus the fiber is rejecting the background light while being well-matched to the coronagraph PSF.  In the imaging case, essentially all light that hits the detector is accepted, meaning that even portions of the uniform background that are very far away from the core of the planet PSF make a measurable contribution.  Note also that the same argument holds for SPLC 2, resulting in a total SNR gain of a factor of $\sim1.9$ in the background-limited case compared to the $10^{-10}$ contrast imaging mode.

\subsection{Sensitivity to Zernike Aberrations}\label{sec:SPLCab}

\begin{figure}
\centering
\includegraphics[scale=0.6]{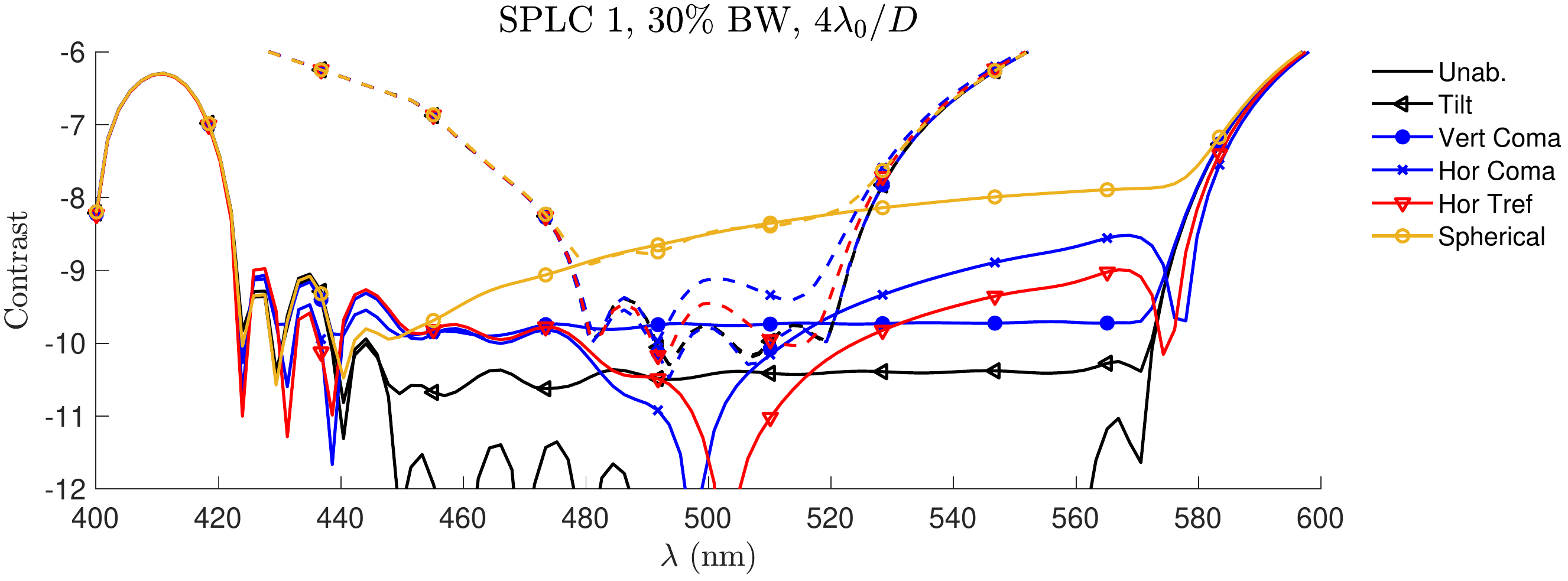}
\includegraphics[scale=0.6]{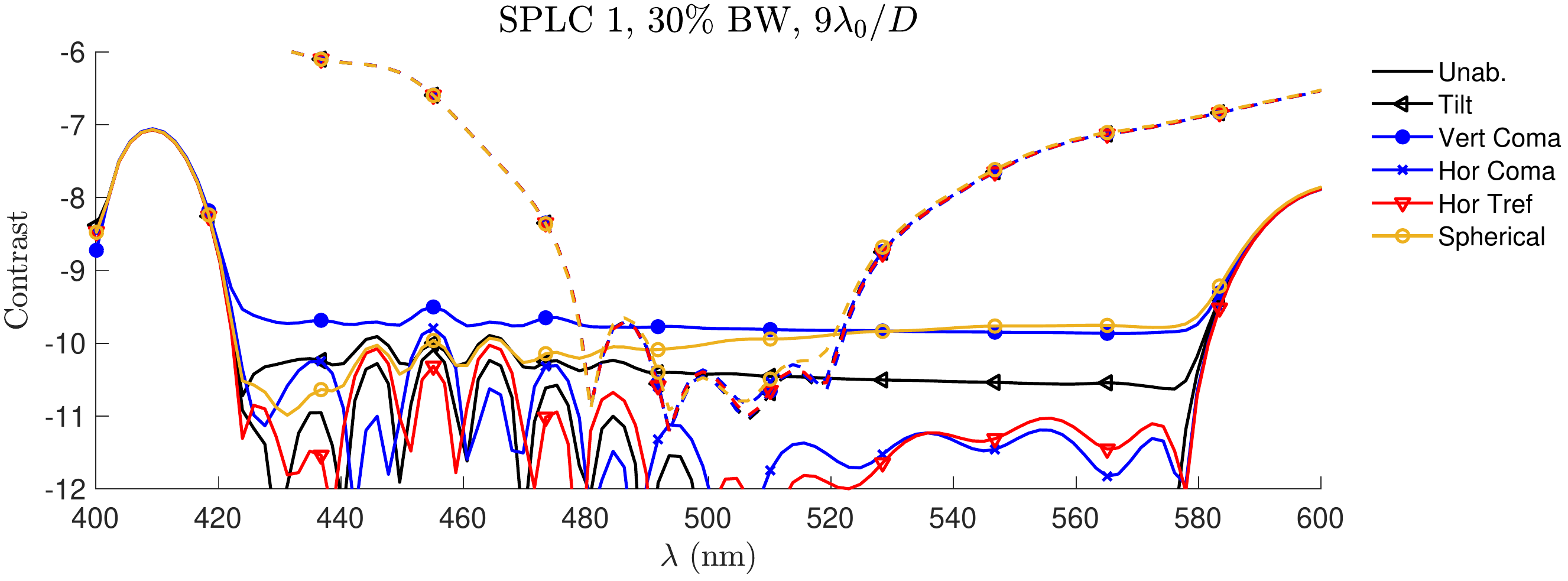}
\caption{Sensitivity of the wavefront control solution at the IWA to 100 pm of low-order Zernikes for SPLC 1.  The top panel shows the comparison at the IWA, while the bottom panel shows the next lenslet out at $9\lambda_0/D$.  The thick black line shows the unaberrated contrast curve, while the colored and markered lines show the effects of the various Zernikes.  Although we evaluated all Zernikes up to Noll index 11 (primary spherical), only the ones with the most significant effects were plotted to enhance plot readability.  The dashed curves are from the imaging solution over a 10\% bandwidth, while the solid curves are from the fiber solution.}  \label{fig:1E10SPLCab}
\end{figure}

\begin{figure}
\centering
\includegraphics[scale=0.6]{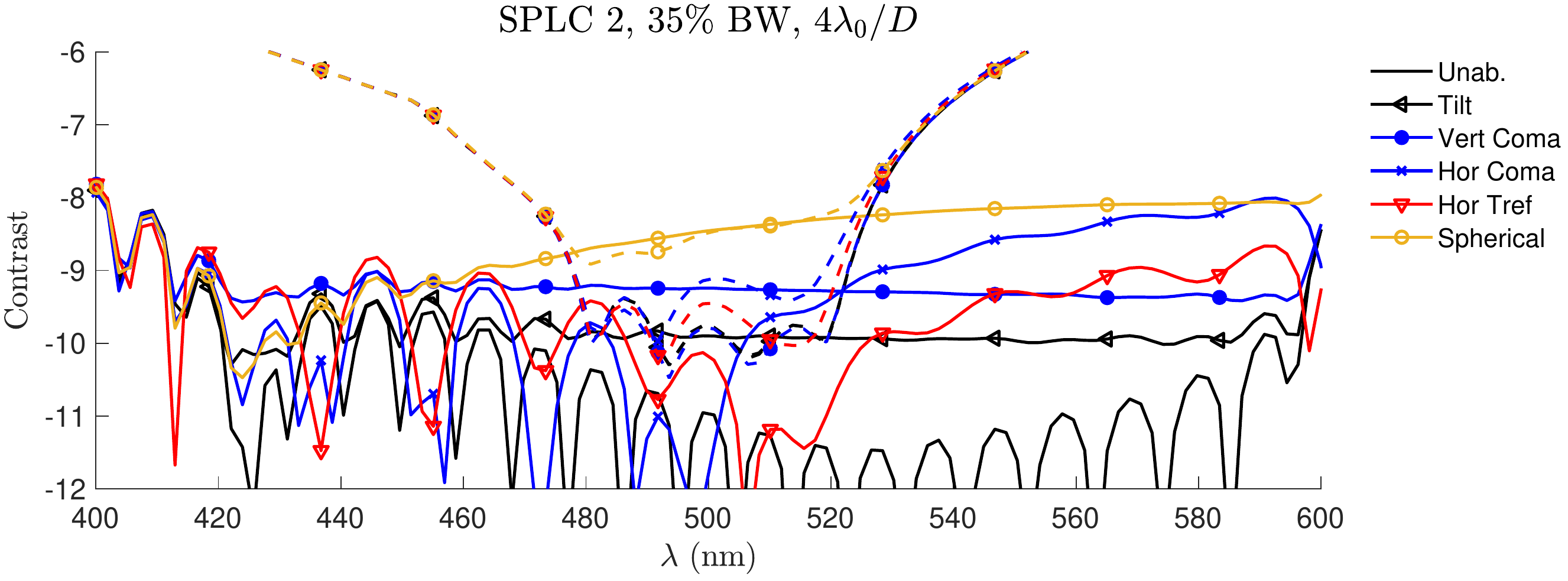}
\includegraphics[scale=0.6]{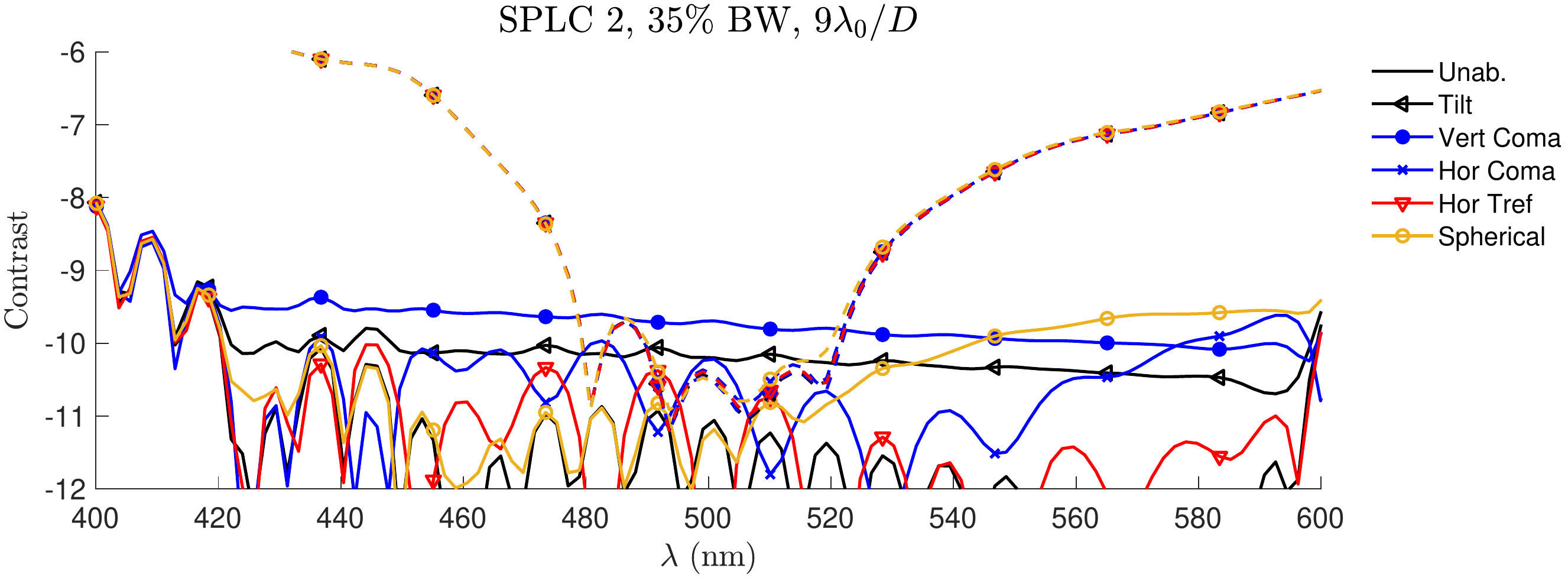}
\caption{Same as Figure~\ref{fig:1E10SPLCab}, but for SPLC 2.  The top panel again shows the IWA lenslet comparison, while the bottom panel shows the next lenslet out at $9\lambda_0/D$.}\label{fig:5E9SPLCab}
\end{figure}

Figures~\ref{fig:1E10SPLCab} and \ref{fig:5E9SPLCab} show the effects of adding 100~pm RMS of individual low-order Zernike aberrations to the fiber EFC and imaging solutions for the SPLC masks.  In each case, the sensitivity to aberrations appears to be manageable, and is not appreciably worse than the sensitivity of the system in imaging mode at the IWA.  This should be expected to be the case, as in a true apples-to-apples comparison where the raw contrast is calculated over the same photometric aperture, a SMF must always display equal or better contrast than an imager.  This is fundamentally due to conservation of energy - fibers can only reject photons, not create them, and a PSF will almost always couple more efficiently than a random speckle field.

\begin{figure}
\centering
\includegraphics[scale=0.6]{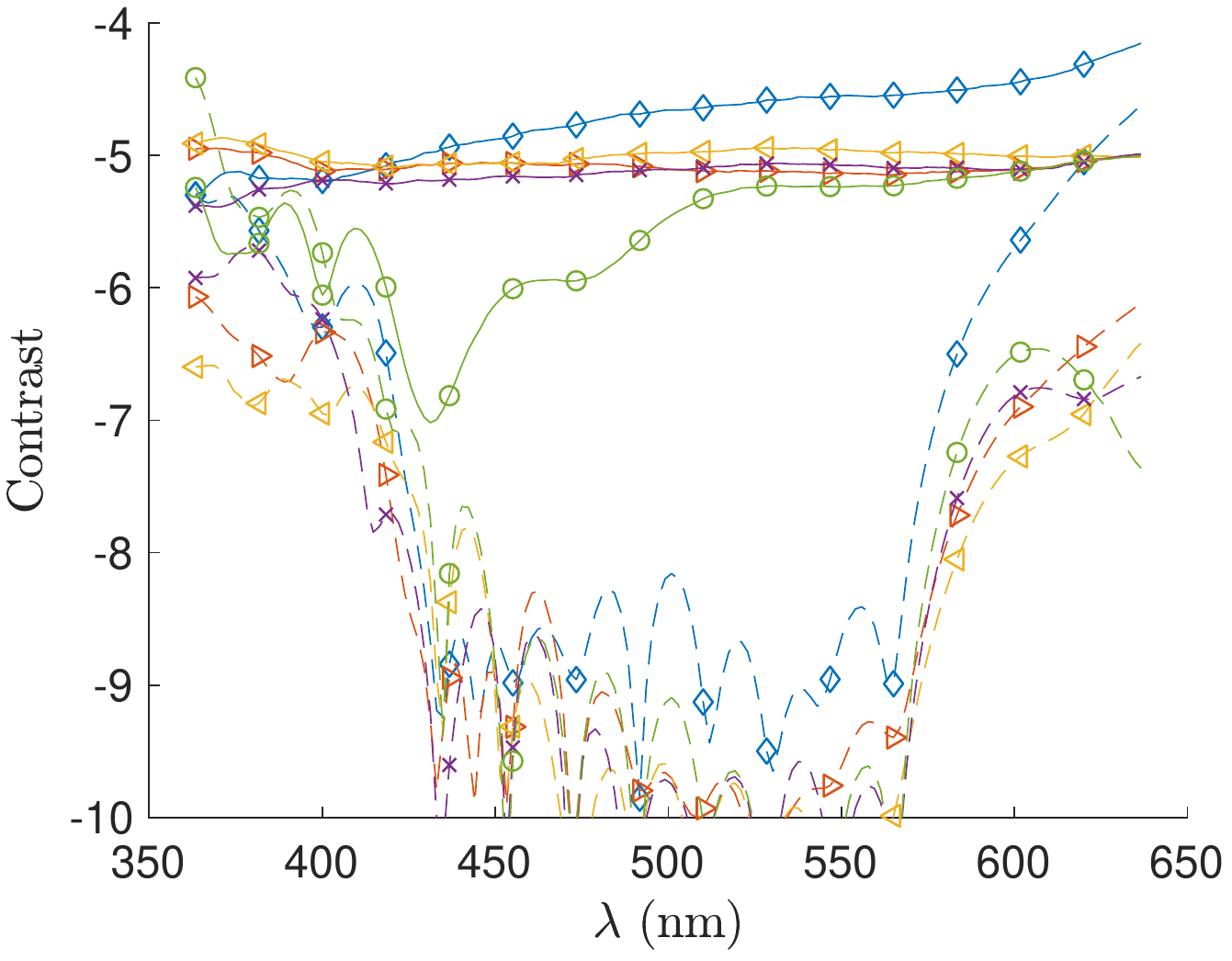}
\caption{Raw contrast vs. wavelength in the lenslet/coronagraph image plane for the 30\% bandwidth fiber and imaging solutions with SPLC 1.  The solid lines show the fiber solution contrast, and the dashed lines the imaging solution contrast.  Each line represents one lenslet; the blue lines show the location of the IWA lenslet.  There is far more light present in the fiber solution, but almost all of it will be filtered out by the fibers.}\label{fig:FibImgRC}
\end{figure}

However, upon moving further out into the field, the aberration sensitivity of the imaging case rapidly drops, while the sensitivity of the fiber solution decreases much more slowly.  The result is that by $9\lambda_0/D$, the imaging solution fares better than the fiber solution when certain aberrations are present.  We believe this to be because a pure fiber EFC solution does not dig a deep dark hole in the coronagraph image plane, so there is much more light that can be shifted into the fiber than in the case of a deep dark hole intended for imaging; Figure~\ref{fig:FibImgRC} compares the raw contrast in the lenslet plane for the 30\% bandwidth SPLC 1 imaging and fiber cases.  Nonetheless, the fiber solution is capable of maintaining its deep contrast performance over a much wider band than the imaging case, and when compared over solutions of the same bandwidth, shows unequivocally better contrast.  Also noteworthy is that SPLC 2's drift sensitivity requirement over time is reduced by 40\% due to the fact that wavefront control exposures can be taken at shorter intervals.

\section{Vortex Coronagraph \& SMF Results}\label{sec:Vortexresults}

\subsection{Maximum Bandwidth and Finite-Sized Star}

\begin{figure}
\centering
\includegraphics[scale=0.56]{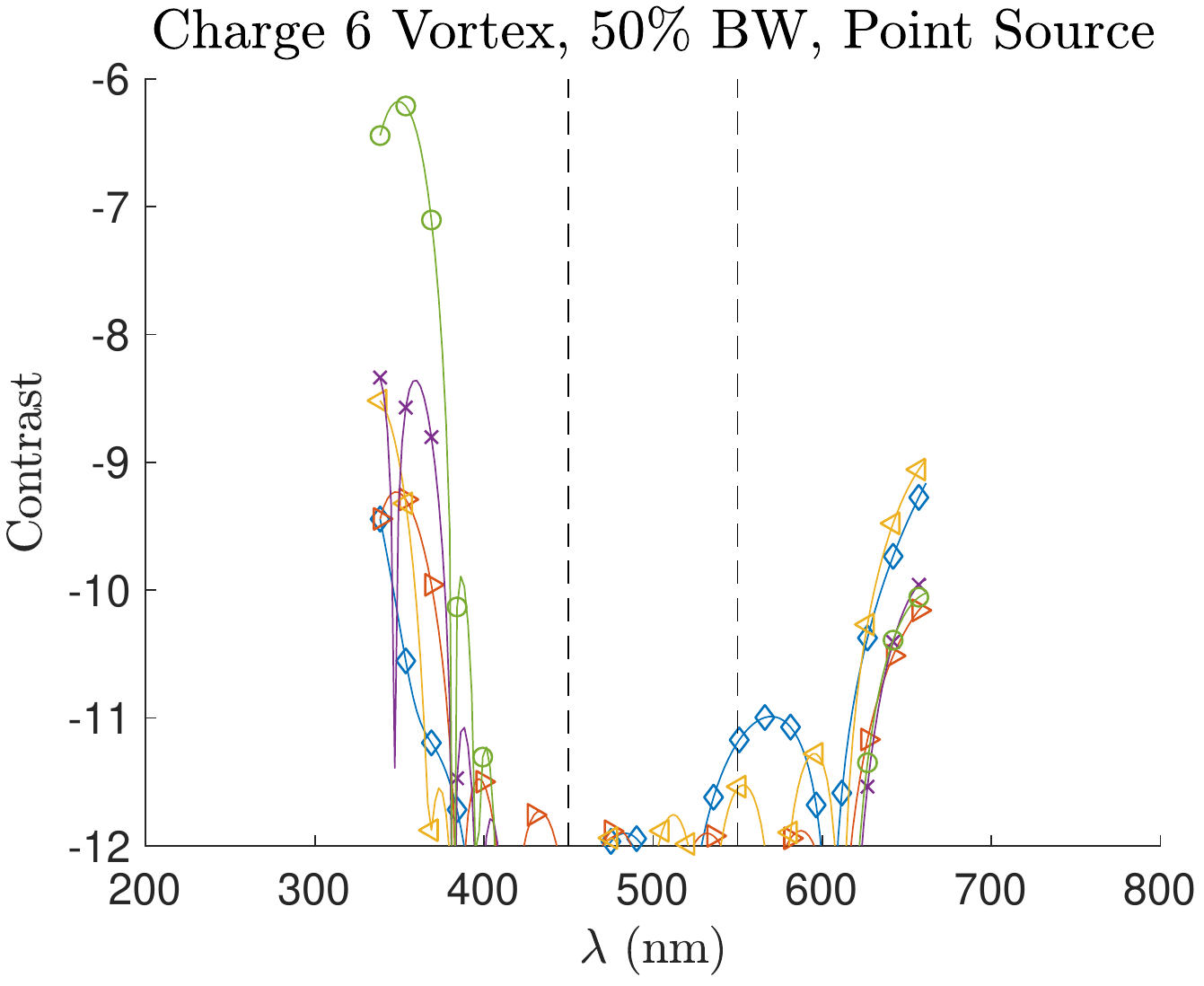}
\includegraphics[scale=0.56]{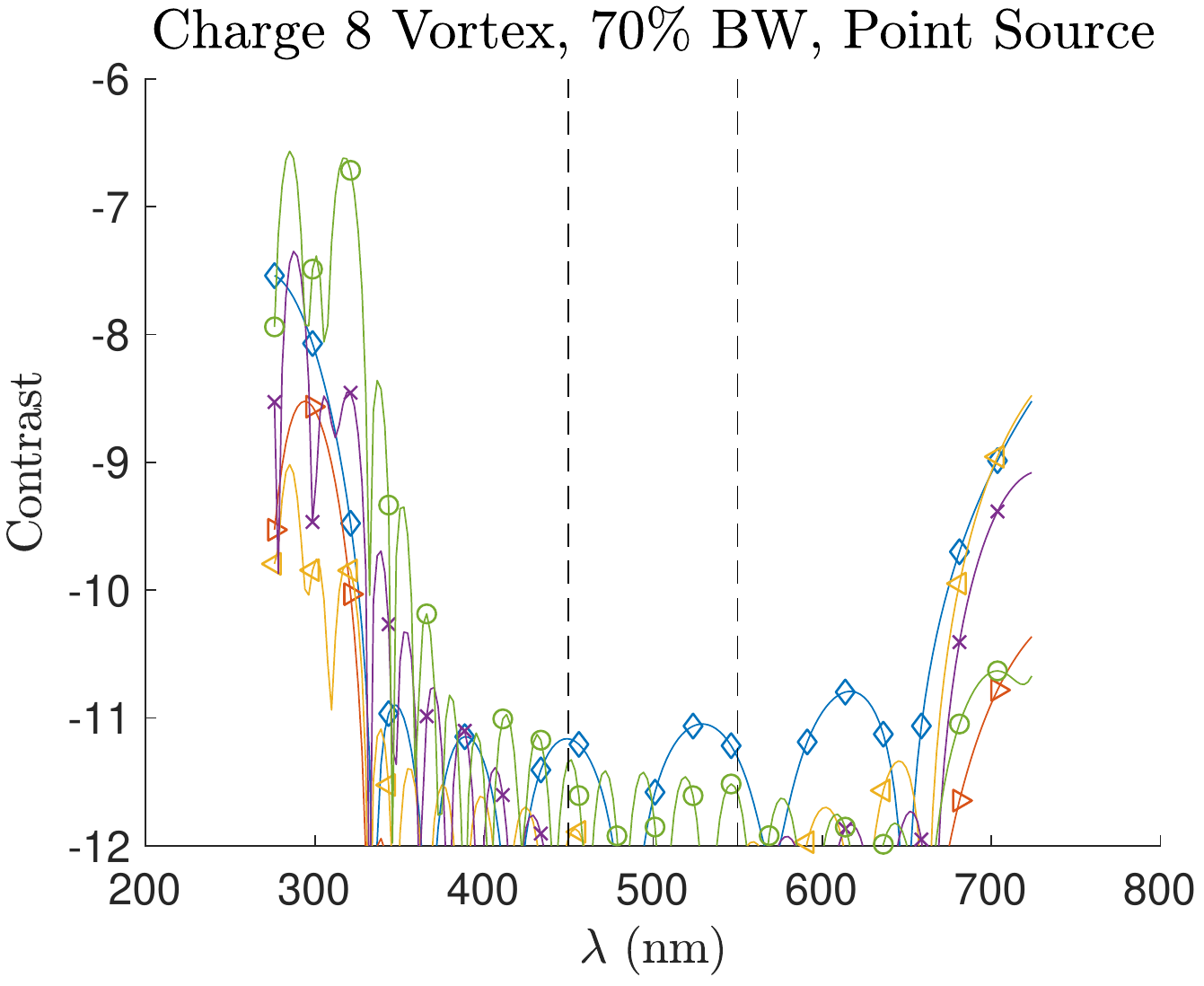}
\includegraphics[scale=0.598]{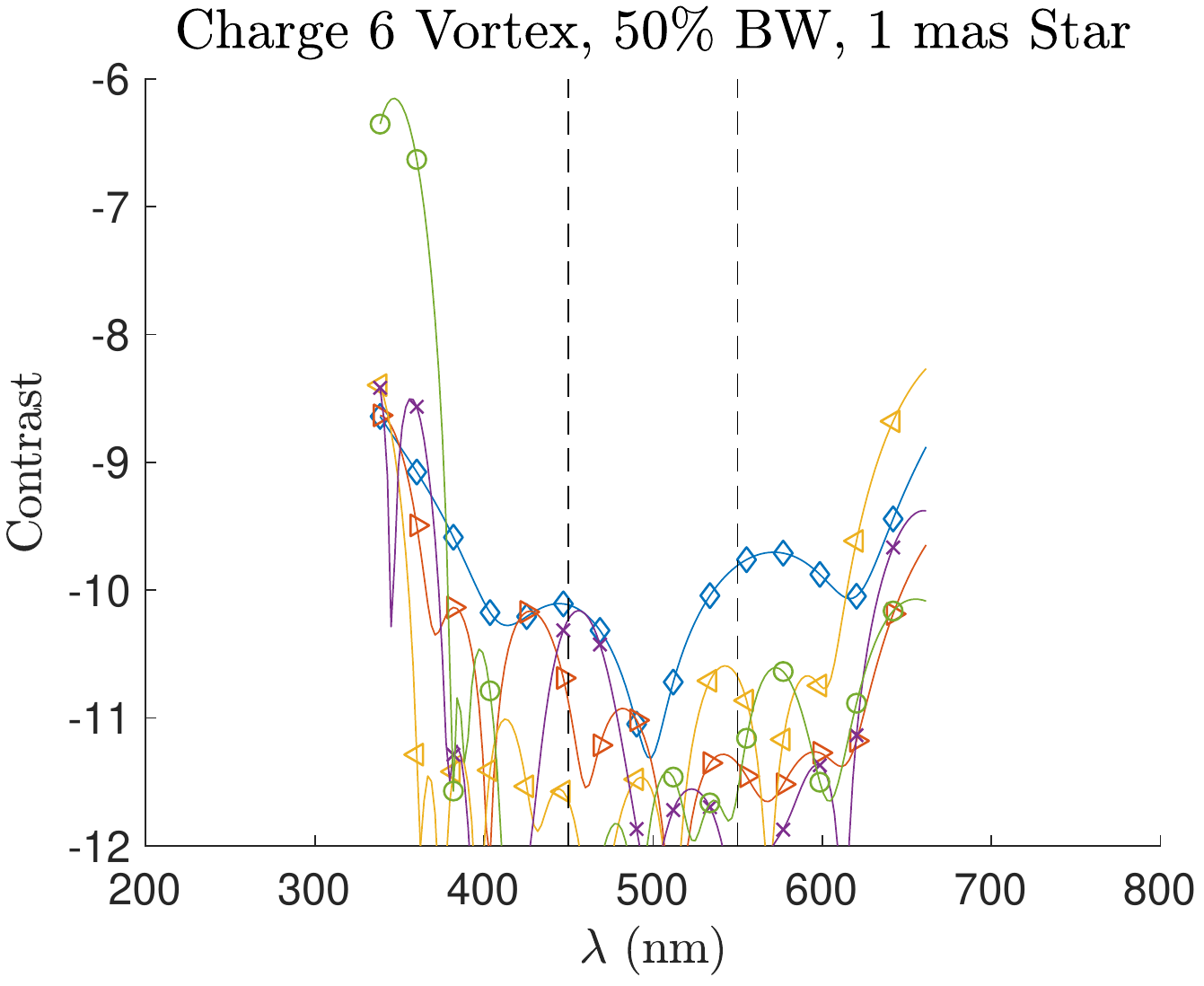}
\includegraphics[scale=0.598]{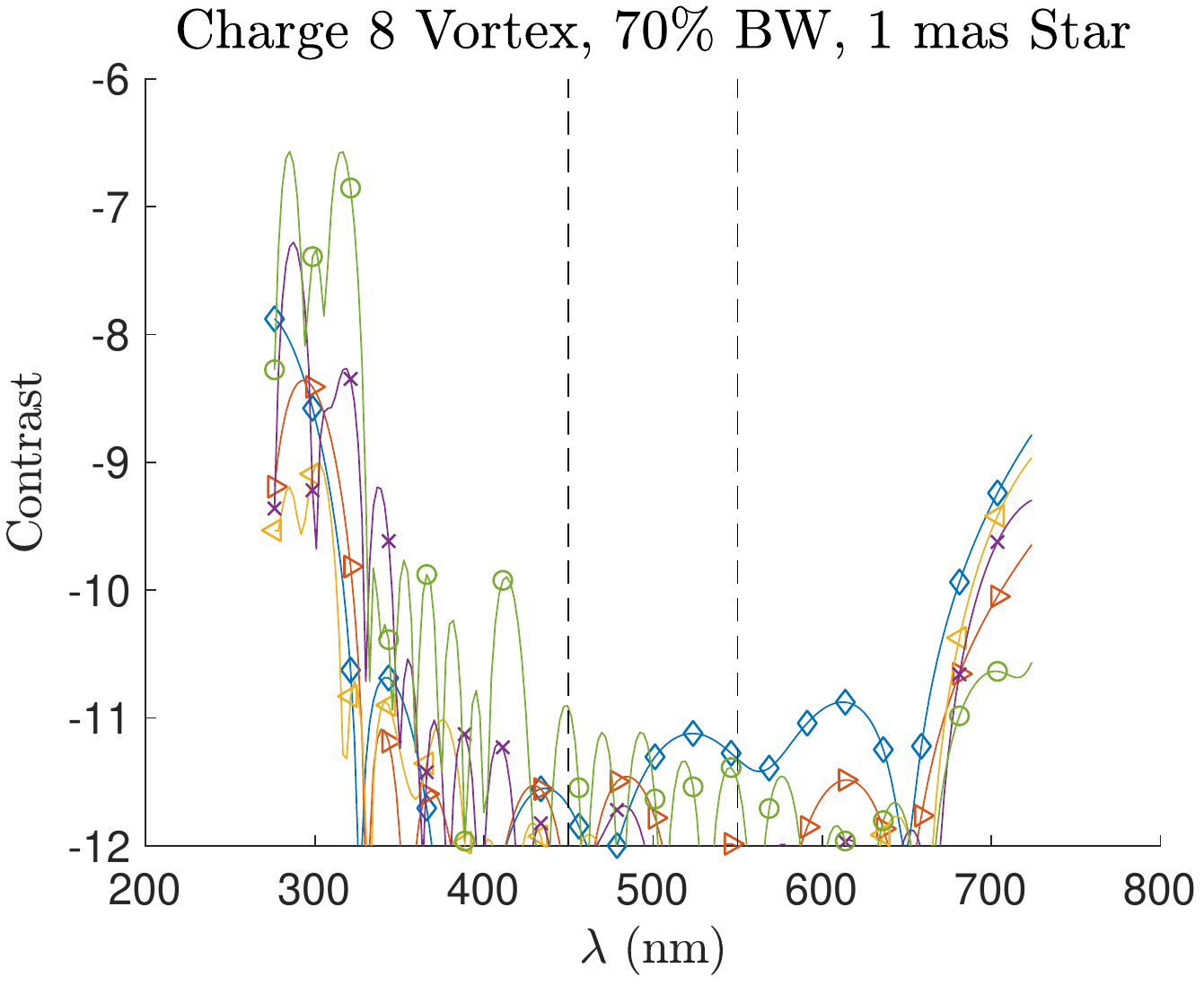}
\caption{Results of simulations testing extended bandwidth for vortex coronagraph designs.  The left column shows the results for the charge 6 mask, while the right column shows the results for the charge 8 mask.  The vertical dashed black lines show the design bandwidth for each mask in imaging mode, while the solid colored lines show the intensity profile with wavelength in each lenslet.  The blue diamond-marked lines correspond to the lenslets placed at the inner working angle.  The top row shows the results for a point source, while the bottom row shows the results using a 1~mas star with tip/tilt sensitivity control enabled.  The purple (x-marked) and green (open circles) lenslets are near the outer edge of the field; in the charge 8 case, the wavelength range is so large that these lenslets lie outside the DMs' correction radius on the blue end, accounting for the failure to maintain good contrast there.}\label{fig:vortexBW}
\end{figure}

Results from the vortex coronagraph are inconclusive with respect to the benefit of using SMFs with regards to bandwidth; however, SMFs show better contrast, lower sensitivity to finite stellar size, and better background-limited SNR than imaging mode.  The top row of Figure~\ref{fig:vortexBW} shows the contrast vs. wavelength for our point source vortex coronagraph simulations.  Using the charge 6 vortex mask, 50\% bandpass is possible with single-mode fibers, similar to the imaging mode, although the fibers do provide up to a factor of $\sim100$ better contrast.  With a charge 8, the usable bandwidth is even larger - at least 70\% is possible; at bandpasses that wide, other factors limit the system's performance more than the mask.  For example, at short wavelengths, the outer lenslets tested cannot maintain the design contrast because they lie outside the DMs' control radius, and the light level rises rapidly.  Therefore, we did not test still wider bandpasses, although they may be possible with a restricted field of view or DMs with higher actuator counts.

The bottom row of Figure~\ref{fig:vortexBW} displays the results of attempting to control for a 1~mas diameter star using the charge 6 and 8 vortex masks on the LUVOIR B pupil.  The results are on the whole encouraging; most fibers remain below or near $10^{-10}$ contrast over the majority of the controlled band, and all display similar or better contrast compared to the finite star imaging mode results.  As expected, the charge 8 mask is able to better control the off-axis starlight.  Given the major similarity of the performance of the fiber and imaging modes with the vortex mask, we cannot say with certainty how much benefit, if any, SMFs will bring to a vortex coronagraph in reality except for the increase in background-limited SNR we will show in Section~\ref{sec:vortexthput}.  That being said, we speculate that using SMFs will enable better correction of chromatic features of the optical system, as they do for the SPLC; this is the subject of future work.

\subsection{Throughput and SNR}\label{sec:vortexthput}

\begin{figure}
\includegraphics[scale=0.6]{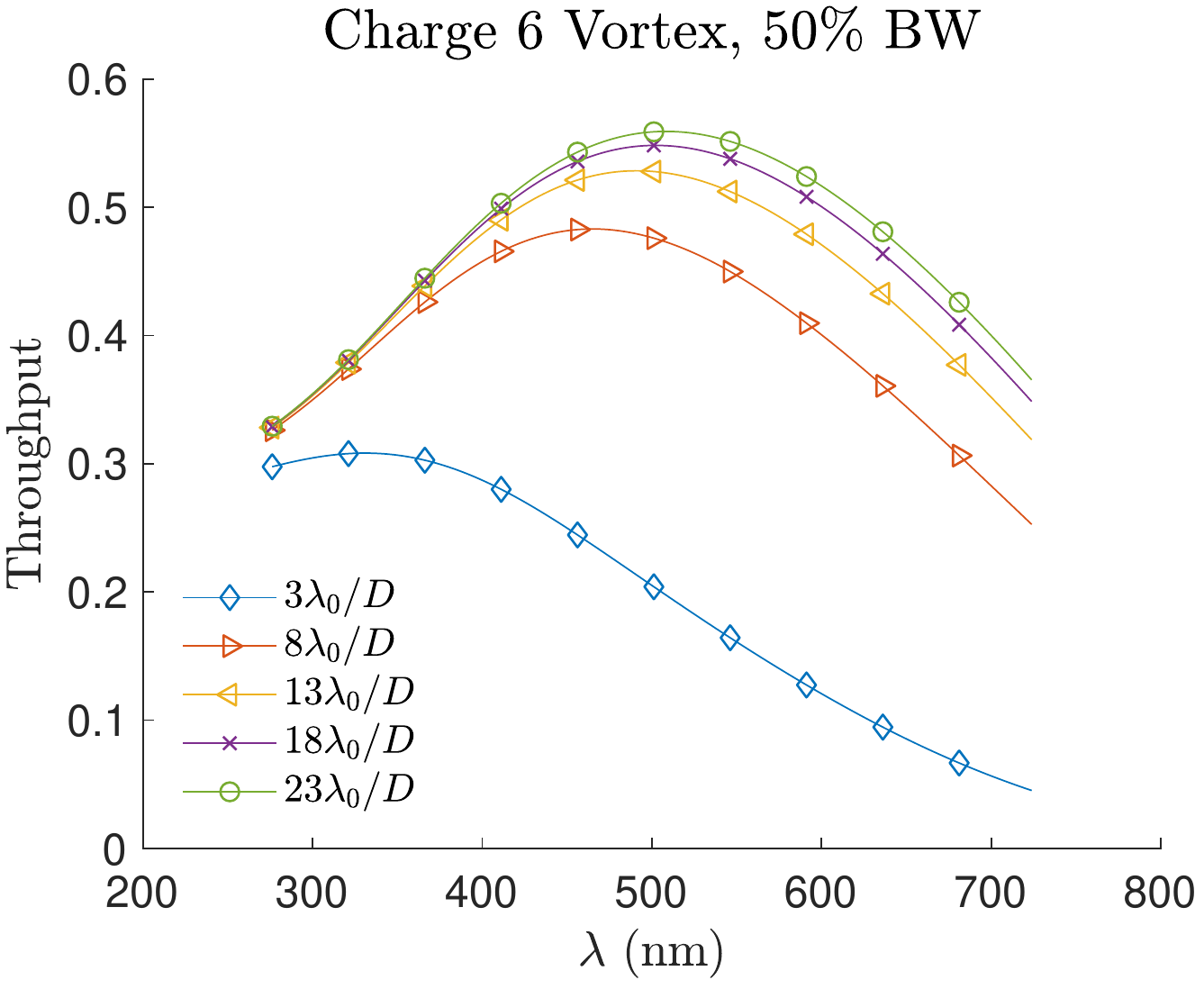}
\includegraphics[scale=0.6]{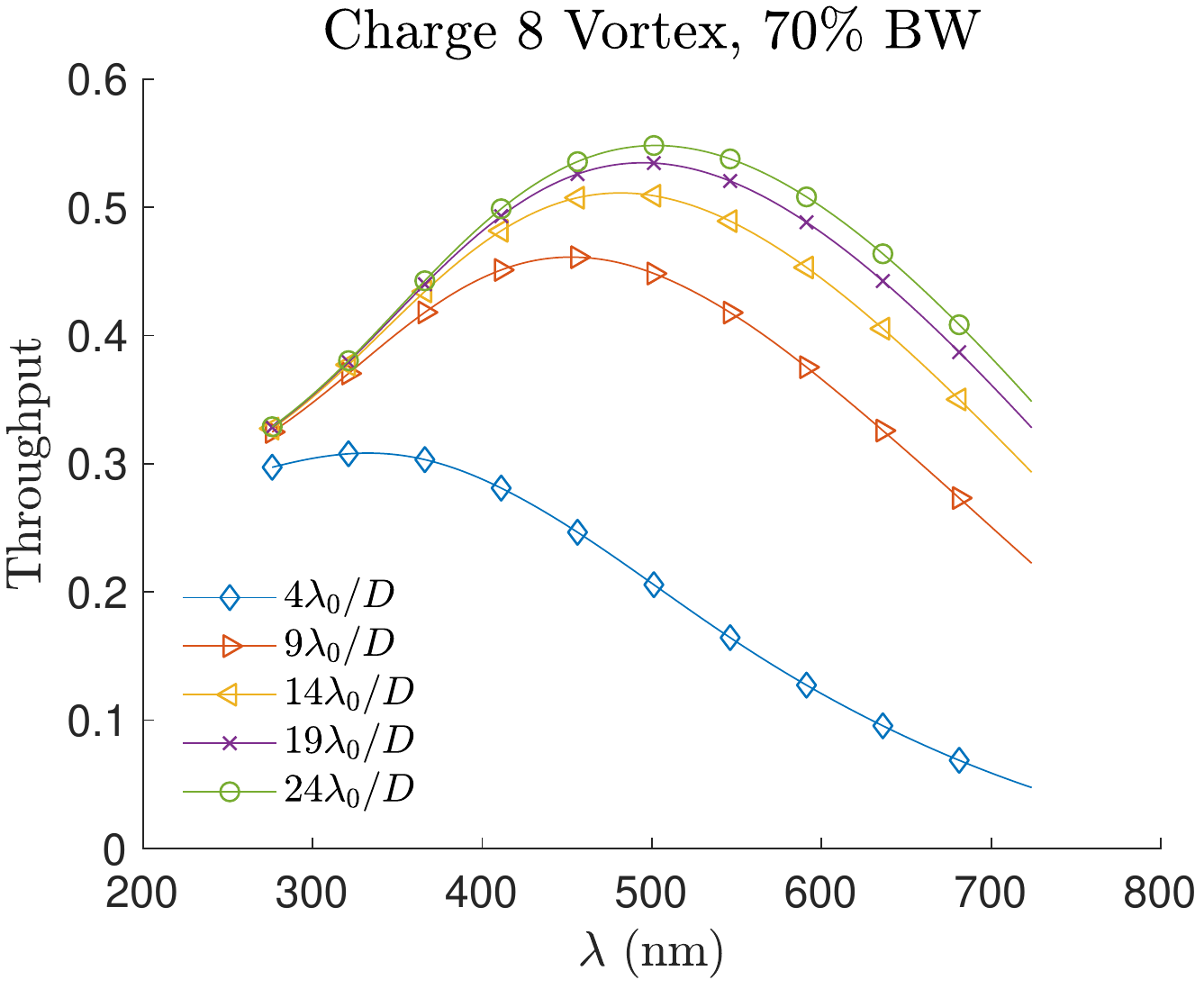}
\caption{Fiber throughput vs. wavelength for the vortex masks.  The left panel shows the results for the charge 6 mask, while the right panel shows the charge 8.  Each line corresponds to one lenslet; the legend gives the lenslet positions in $\lambda_0/D$.  The lenslet spacing for each mask is $5\lambda_0/D$.}\label{fig:vortexthput}
\end{figure}

Figure~\ref{fig:vortexthput} shows the fiber throughput vs. wavelength for the charge 6 and 8 vortex masks; throughput gains are in the line with those seen in the SPLC designs due to the change in throughput definition.  To estimate the background-limited SNR from an infinite uniform background, we followed a similar procedure to the one we used for the SPLC masks with the fiber.  Because no easy analytical argument is possible when determining the throughput of a uniform background in the imaging case due to the vortex's non-uniform throughput curve, we instead simulated a densely-packed array of point sources both inside and surrounding a FWHM-sized aperture centered on the lenslet at $13\lambda_0/D$ for the charge 6 mask.  The point sources were arranged in a square $4.88\lambda_0/D$ on a side, space $0.04\lambda/D$ apart. The simulations showed that the average background throughput in imaging mode is 100\%, while with the fiber, $T_{\textrm{pl}} = 0.49$ and $T_{\textrm{bg}} = 0.18$, resulting in a relative SNR increase of a factor of $\sim1.9$ from the background-limited imaging case.  The charge 8 shows slightly larger gains, reaching a factor of $\sim2.1$ uniform-background-limited SNR gain from switching to the SMF.

\subsection{Sensitivity to Zernike Aberrations}\label{sec:vortexab}

\begin{figure}
\centering
\includegraphics[scale=0.6]{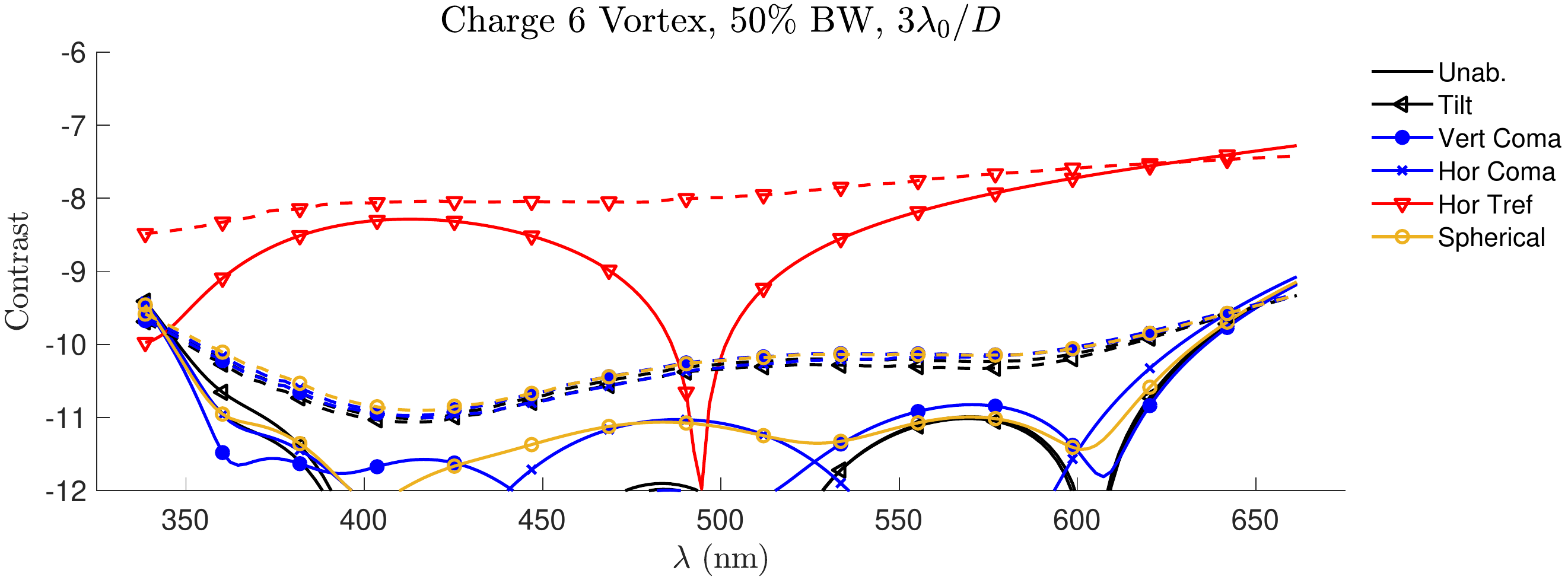}
\includegraphics[scale=0.6]{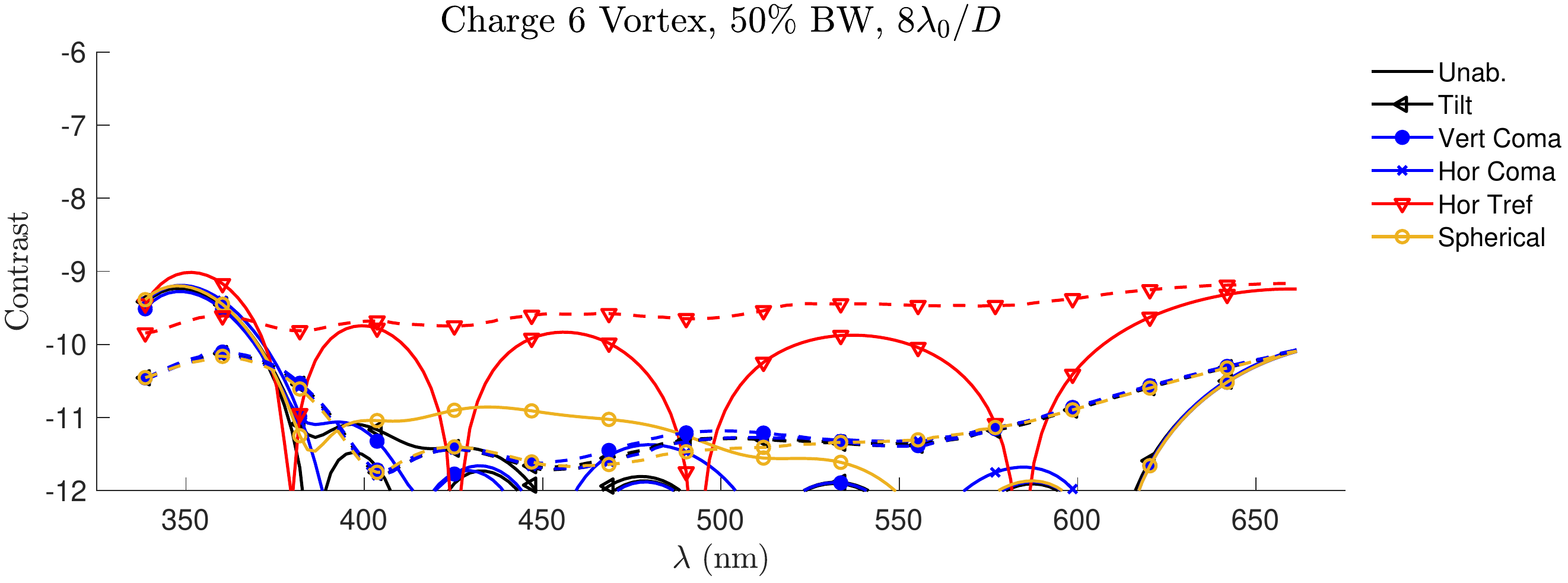}
\caption{Aberration sensitivities of the charge 6 vortex mask.  100 pm RMS of each low-order Zernike aberration was introduced and the resulting contrast curves plotted.  The top panel shows the results at the IWA, while the bottom panel shows the results at the next lenslet out, at $8\lambda_0/D$.  As with the SPLC figures, the dashed lines show the results from imaging mode, while the solid lines show the fiber solution.  Trefoil is the only tested aberration which is transmitted well by the mask.}\label{fig:C6IWAab}
\end{figure}

\begin{figure}
\centering
\includegraphics[scale=0.6]{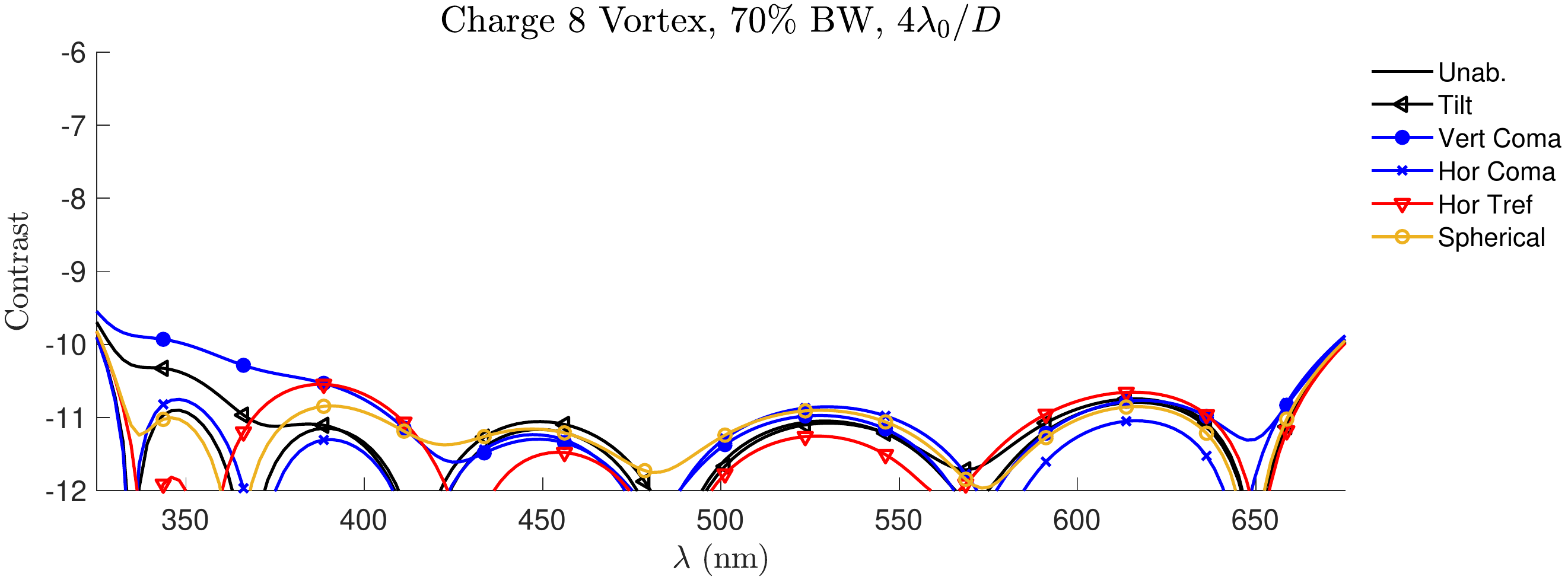}
\includegraphics[scale=0.6]{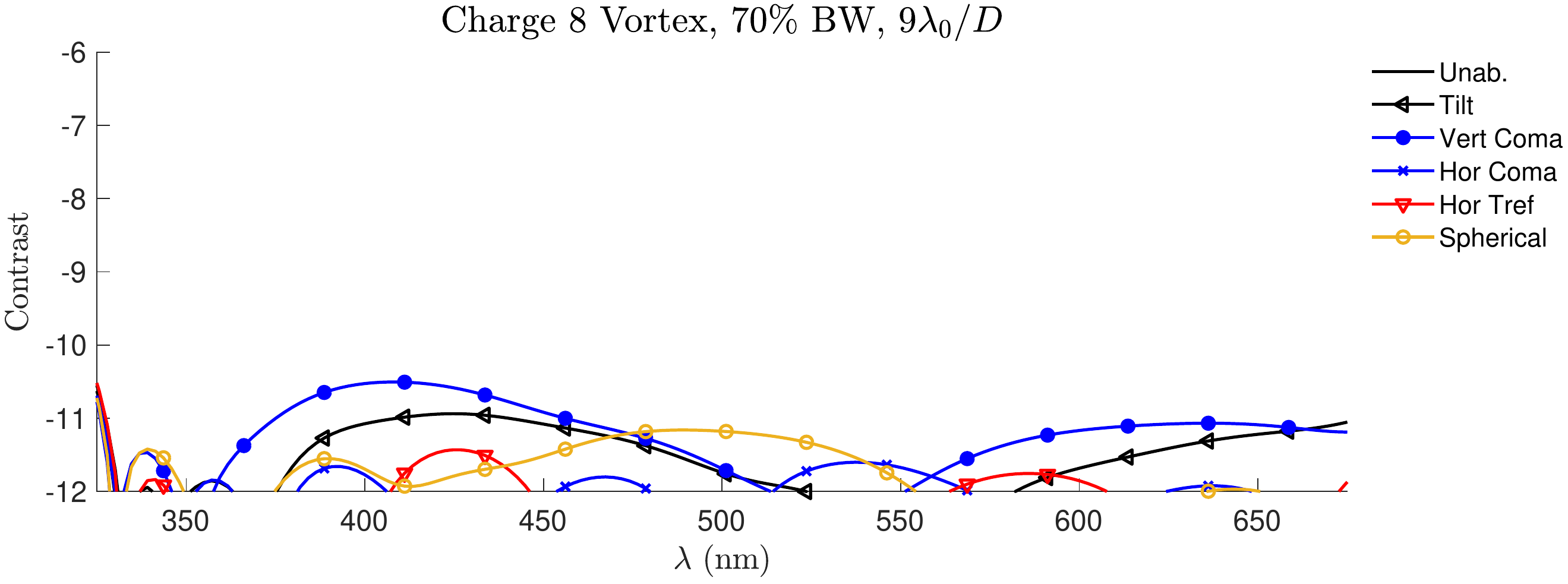}
\caption{Same as Figure~\ref{fig:C6IWAab}, but for the charge 8 mask, and with no imaging mode results plotted.  None of the low-order Zernikes are transmitted well by the mask, showing the charge 8's excellent aberration suppression.}\label{fig:C8IWAab}
\end{figure}

Vortex masks on unobscured apertures are known to be good low-order aberration filters, a result that remains true with SMF imaging.  Figures~\ref{fig:C6IWAab} and \ref{fig:C8IWAab} show the sensitivity to low-order Zernike aberrations of the charge 6 and 8 vortex masks, respectively.  The results are as would be expected of vortex masks - the charge 6 is very sensitive to trefoil, and insensitive to most lower-order Zernike aberrations, while the charge 8 is insensitive to all low-order Zernike aberrations tested.  The charge 6's reaction to trefoil is not as severe further out in the field of view, making that issue not as dire as it first appears.

\section{Discussion \& Conclusions}\label{sec:conclusion}

\begin{figure}
\centering
\includegraphics[scale=0.58]{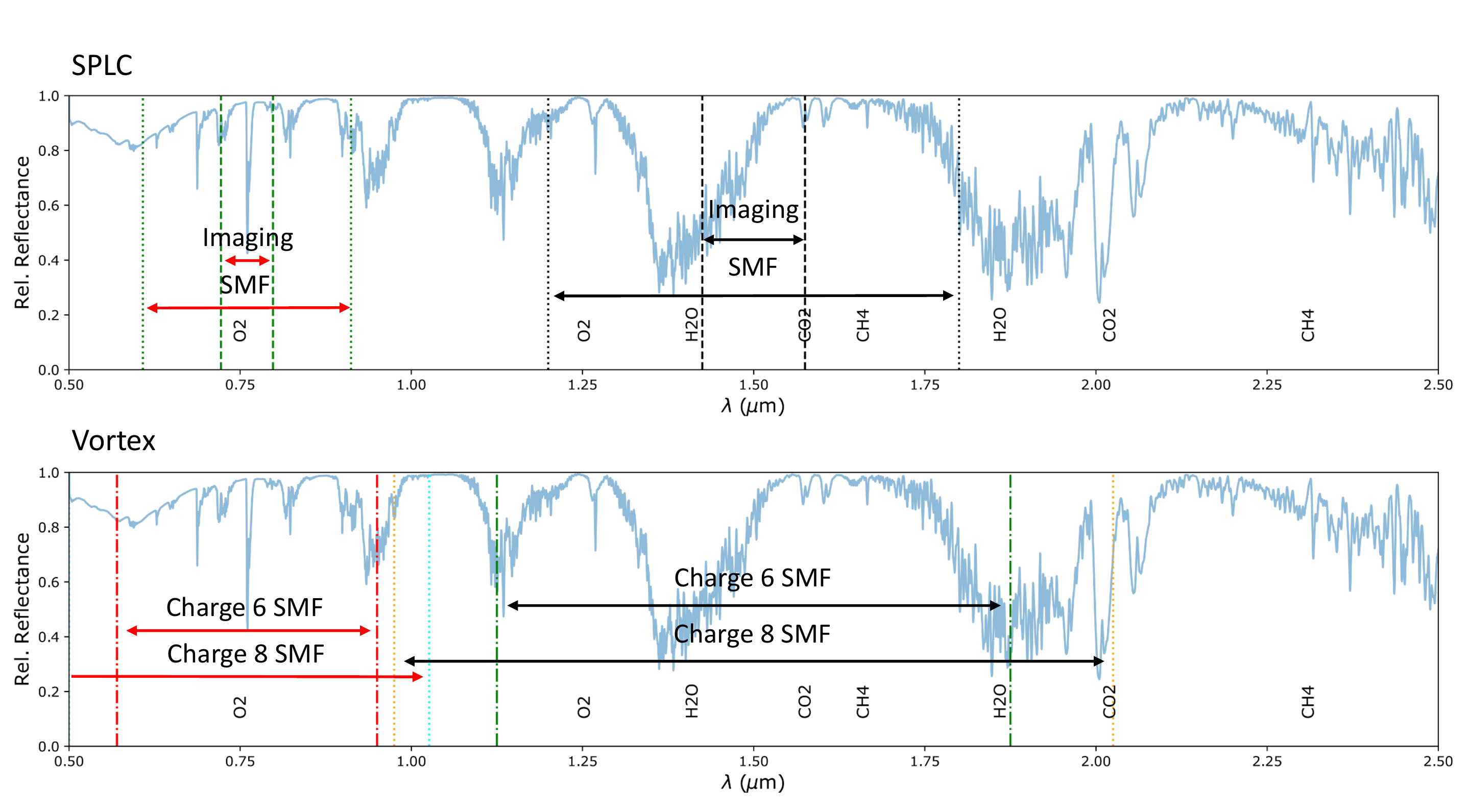}
\caption{Simulated $\lambda/\Delta\lambda=1,000$ spectrum of an Earth-like planet from Ref.~\citenum{Coker18a} showing the relative bandwidths for the different coronagraph masks simulated.  Major spectral lines/line complexes are labeled; most of the unlabeled lines are due to water.  The top panel is for the SPLC masks, while the bottom panels shows the vortex masks.  Because SPLC 2 offers strictly better peformance than SPLC 1 except on larger stars, SPLC 1 is not plotted here.  For each mask type, we show a comparison for two bands, one centered on the oxygen A line at 760~nm, and another centered in the near-infrared at 1.5~$\mu$m.  For the SPLC, the imager or conventional spectrograph can only observe the immediate vicinity of oxygen A, while the SMF-fed spectrograph would be able to observe the other nearby oxygen and water lines on the same exposure.  If a charge 8 vortex is used, the spectrograph would be able to simultaneously observe from 1-2~$\mu$m in a single exposure, absent other optical limitations.  In both the SPLC and vortex cases, the SMF-fed spectrograph would be able to capture water, oxygen, methane, and carbon dioxide in a single exposure, provided a long enough integration time.}\label{fig:EBspectrum}
\end{figure}

The implications of comparing the fiber and imaging results for on-axis telescopes are stark.  Table~\ref{table:coroSNRgain} gives a summary of our findings; through the use of SMFs, the bandwidth of the coronagraph may be extended by up to a factor of 3.5 while improving throughput/SNR.  Figure~\ref{fig:EBspectrum} shows the difference in the simultaneously coverable bands between SMFs and imaging mode on the model spectrum of an Earth-like planet.  When observing in the near-infrared, an SPLC feeding into a SMF is able to observe water, oxygen, carbon dioxide, and methane simultaneously, while the imaging mode or conventional spectrograph requires tuning to at least three separate bandpasses to achieve the same.  In the visible, the SMF-fed spectrograph can observe not just the oxygen line at 760~nm, but also the surrounding oxygen and water lines.  For the achromatic vortex mask, even more lines are observable; a SMF-fed spectrograph behind an achromatic vortex could, for example, observe the entire visible spectrum at once, or cover the J and H bands with room left over.  Either can simultaneously capture water, oxygen, carbon dioxide, and methane features.

\begin{table}
\centering
\caption{Coronagraph Efficiency}
\begin{tabulary}{\textwidth}{C|C|C|C|C|C|C|C}
\hline
\hline
Coronagraph Mask & $T_{pl}$ & $T_{bg}$ & Background-Limited SNR Boost & Bandwidth \\ \hline
SPLC 1 Img. & 0.12 & 0.41 & - & 10\% \\ \hline
SPLC 1 Fiber & 0.19 & 0.078 & 1.5 & 30\% \\ \hline
SPLC 2 Fiber & 0.28 & 0.094 & 1.9 & 35\% \\ \hline
Charge 6 Vortex Img. & 0.25 & 1 & - & 50\% \\ \hline
Charge 6 Vortex Fiber & 0.49 & 0.18 & 1.9 & 50\% \\ \hline
Charge 8 Vortex Img. & 0.22 & 1 & - & 70\% \\ \hline
Charge 8 Vortex Fiber & 0.47 & 0.176 & 2.1 & 70\% \\
\hline
\end{tabulary}
\label{table:coroSNRgain}
\end{table}

\begin{figure}
\centering
\includegraphics[scale=0.6]{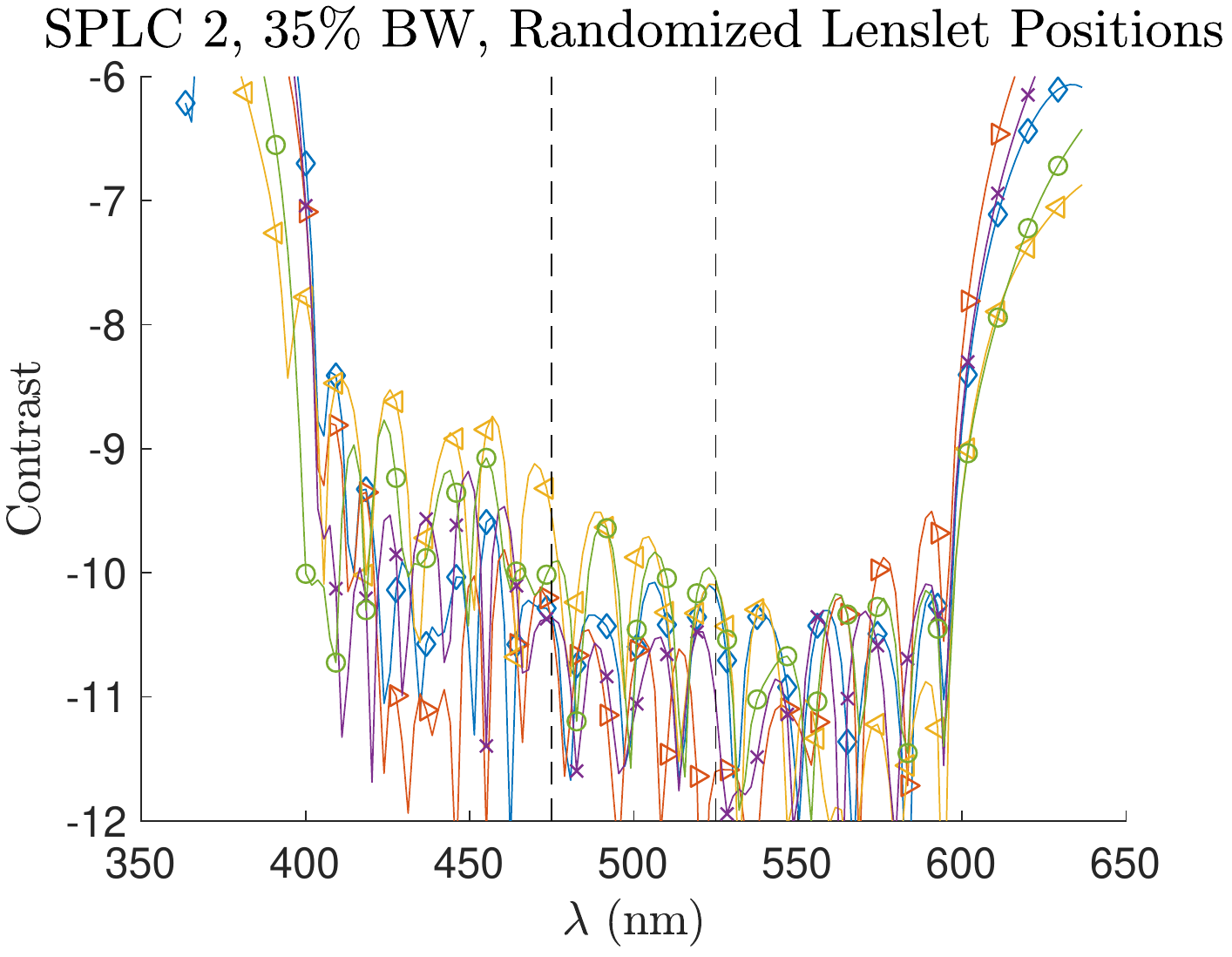}
\caption{Base-10 log raw contrast vs. wavelength for the randomized lenslet trial using SPLC 2.  Each line represents one lenslet.  The contrast performance is a factor of $\sim10$ worse than the case with all the fibers in a line, suggesting that lining up the lenslets makes them easier to control via fiber EFC, although performance is still good, and far better than it would be in imaging mode.}\label{fig:randomcontrast}
\end{figure}

One possible objection to our work is that we simulated evenly spaced lines of lenslets rather than random positions, such that the DM does not have to reproduce as many spatial frequencies and allowing it find a better wavefront control solution.  We therefore also generated a list of five random lenslet angular separations and position angles and attempted to control them using SPLC 2; Table~\ref{table:randlenspos} contains the generated lenslet positions.  Figure~\ref{fig:randomcontrast} shows the raw contrast vs. wavelength for those lenslets.  This is the same trial as was shown in Section~\ref{sec:WFC}.  The contrast performance is a factor of $~\sim10$ worse than the case where all the lenslets are in a line, although most of the lenslets are able to maintain near or better than $10^{-10}$ contrast over large portions of the bandwidth.  This indicates that the results obtained in Sections~\ref{sec:SPLCresults} and \ref{sec:Vortexresults} are achievable with the lenslets in arbitrary positions within the coronagraph's field of view.

\begin{table}
\centering
\caption{Randomized Lenslet Positions}
\begin{tabulary}{\textwidth}{C|C}
\hline
\hline
Angular Separation $(\lambda_0/D)$ & Position Angle ($^{\circ}$) \\ \hline
23.59 & 80.83 \\ \hline
22.87 & 30.37 \\ \hline
5.49 & 147.80 \\ \hline
15.38 & 143.89 \\ \hline
7.22 & 53.31 \\
\hline
\end{tabulary}
\label{table:randlenspos}
\end{table}

A further extension of this work would be to turn the five-fiber concept presented here into a discovery instrument with many fibers covering the entire field of view behind a microlens array - indeed, such an instrument is the first design we tested.  Unfortunately, it is not practical.  When 120 fibers are used in a five-ring hex-packed arrangement covering a region from $4-15\lambda_0/D$, extended bandwidth is not possible while maintaining $10^{-10}$ contrast, and the system contrast performance approximates that of a coronagraph used in standard imaging mode.  Further, the rejection capabilities of the fibers mean that when the lenslets are sized for optimal on-axis throughput, the throughput for targets on the edges of lenslets drops to effectively zero, meaning that a three-position dither pattern is needed to fill in the gaps in between lenslets.  Reducing the lenslet radii raises the throughput at the lenslet edges at the cost of on-axis throughput; reducing the radii enough to negate the need for dithering ultimately reduces the throughput too much to be competitive with a traditional multimode-fiber-fed spectrograph.

Alternatively, one may ask whether there are further performance benefits to be had by using fewer than five fibers.  Our initial bandwidth testing was performed using one fiber; we found that one fiber could sustain up to $10-20\%$ relatively higher bandwidth than five fibers (e.g., $\sim35\%$ for one fiber vs. $30\%$ for five), but there were no appreciable gains in ultimate achievable contrast or in throughput.  This suggests that the maximum achievable bandwidth is a weak function of the number of the fibers being controlled simultaneously, such that using more than five fibers in a configurable multi-object instrument is feasible to some degree, even though a discovery instrument is impractical.

Other efforts to use SMFs with coronagraphs are underway.  The SCAR coronagraph\cite{SCAR1,SCAR2} uses a pupil-plane phase plate to mode-shape the incoming light before focusing it onto a lenslet array backed by single-mode fibers, achieving moderate contrast ($10^{-4} - 10^{-5}$) and high throughput over a 10-20\% bandpass.  There are two major differences with our work:  first, that the SCAR design is for ground-based instruments, so that very high contrast ($\sim10^{-10}$) is not expected (or indeed possible) and second, that we use the DMs to shape the light for the fibers instead of a static optic.  Our work is simply in a different regime than theirs, sacrificing throughput for a large gain in contrast to enable observations of Earth-like exoplanets.

Meanwhile, Ref.~\citenum{Mawet17} used speckle nulling on a single SMF to achieve a starlight suppression gain of a factor of $500-1000$ in monochromatic light in the lab, along with performing broader band simulations up to 10\% bandwidth which demonstrated a similar suppression gain beginning with a raw coronagraph contrast of $10^{-3}$.  They also placed their lone fiber directly in the focal plane.  In our work, we used lenslets in the focal plane instead because a microlens array enables denser sampling of the image plane when using multiple fibers.  To obtain good coupling into the fiber, the size of the incident PSF must be roughly the same size as the mode field diameter, which is slightly larger than the fiber core.  For a typical SMF, the mode field diameter is $\sim8\,\mu$m, while the cladding diameter is $\sim125\,\mu$m.  This means that, assuming a mode field width of $3.2\lambda_0/D$, the closest possible core-to-core fiber spacing is $\sim50\lambda_0/D$, rendering it virtually impossible to observe multiple planets in a system simultaneously.  Even using a multicore SMF does not provide the small effective core spacing that lenslets can of just $3.2\lambda_0/D$.  Nor does a multicore fiber offer the same positioning flexibility of individual lenslets, where a system can be constructed to allow nearly arbitrary lenslet positions.

Ref.~\citenum{Sayson18} and \citenum{Sayson19} were the first to use EFC through a SMF and demonstrate its effectiveness in the lab.  We build on that work here by controlling multiple fibers over a much broader spectral band than they did.  Their simulation results using their wavefront control code also show similar tip/tilt sensitivity to ours, which is very encouraging.

In this paper, we have presented the results of our simulations of the performance of single-mode fibers paired with standard coronagraphic masks, with a focus on future space-based telescope concepts such as HabEx and LUVOIR.  We find that using single-mode fibers improves usable bandwidth by a factor of 2.5-3.5 for the SPLC (and we would expect a similar improvement for the case of a vector vortex whose characteristics would otherwise limit its bandwidth to $10 -20\%$), background-limited SNR by up to a factor of $\sim2$, with little to no increase in the sensitivity of the system to low-order aberrations near the IWA.  We find that SMFs enable the use of coronagraph masks with poorer raw contrast, allowing further increases in throughput and spectral bandwidth.  We additionally find that much of this improvement is due to the use of single-mode fibers, rather than other sources such as making a smaller dark hole.  Ultimately, we find that the total integration time of spectral coronagraphic observations may be reduced by a factor of $\sim8$ when attempting to cover a wide spectral range using a SPLC with an on-axis telescope.  Further improvement may well be possible through the use of more carefully optimized lenslets and coronagraph masks.  Given these huge apparent benefits, single-mode fibers should be an integral part of any future attempt to spectrally characterize Earth-like exoplanets.

\acknowledgments

This work was supported by an appointment to the NASA Postdoctoral Program at the Jet Propulsion Laboratory, California Institute of Technology, administered by the Universities Space Research Association under contract with NASA.  This research was carried out at the Jet Propulsion Laboratory, California Institute of Technology, under a contract with the National Aeronautics and Space Administration.  Copyright 2019 California Institute of Technology. Government sponsorship acknowledged. All rights reserved.

\bibliography{cokerbib}
\bibliographystyle{spiejour}

\vspace{2ex}\noindent\textbf{Carl Coker} is a NASA Postdoctoral Program Fellow at the Jet Propulsion Laboratory. He received his PhD in astronomy from the Ohio State University in 2017.  He is the author of 4 refereed journal papers. His current research interests include combining high-contrast coronagraphy and high-resolution spectroscopy, and enabling the observation of Earth-like exoplanets. He is a member of SPIE.

\vspace{1ex}
\noindent Biographies and photographs of the other authors are not available.

\listoffigures
\listoftables

\end{spacing}
\end{document}